\documentclass[11pt]{article}
\usepackage{float}
\usepackage[margin=1in]{geometry}
\usepackage{amsmath,amssymb,amsthm}
\usepackage{graphicx}
\usepackage{float}
\usepackage{booktabs}
\usepackage{natbib}
\usepackage{microtype}
\usepackage{authblk}
\usepackage{etoolbox}
\usepackage{anyfontsize}

\usepackage[
    colorlinks=true,
    linkcolor=cyan,
    citecolor=cyan,
    urlcolor=cyan
]{hyperref}

\newtheorem{theorem}{Theorem}
\newtheorem{proposition}{Proposition}
\newtheorem{corollary}{Corollary}

\newtheorem{lemma}{Lemma}

\theoremstyle{definition}
\newtheorem{condition}{Condition}
\newtheorem{definition}{Definition}
\usepackage{rotating}
\usepackage{placeins}
\usepackage{caption}
\usepackage{float}
\newcommand{\OW}{\mathrm{OW}}
\newcommand{\CO}{\mathrm{CO}}

\makeatletter
\patchcmd{\@maketitle}
  {\LARGE \@title}
  {\fontsize{16}{19.2}\selectfont\@title}
  {}{}
\makeatother

\title{\textbf{Limited Follow Up Overlap in Causal Survival Analysis: Censoring-Adjusted Weighting for Restricted Mean Survival Time}}

\author[1]{Nicholas Tenkorang}
\author{Emmanuella Duah}
\author{Bismark Nyarko}

\affil[1]{Department of Mathematical Sciences,
University of Texas at El Paso}

\date{}

\begin{document}

\maketitle

\noindent\rule{\textwidth}{0.8pt}

\begin{abstract}
Overlap weighting addresses limited treatment overlap by targeting individuals for whom treatment assignment is uncertain. With right censored outcomes, however, adequate treatment overlap does not ensure adequate follow up. We show that the overlap weighted restricted mean survival time difference can have an infinite semiparametric efficiency bound even under randomized treatment and pointwise censoring positivity. We propose censoring adjusted overlap (CO) weighting, that accounts jointly for treatment assignment and the probability of remaining under observation through the restriction horizon. CO approaches ordinary overlap weighting as the strength of the follow up adjustment decreases and progressively downweights covariate profiles with weak follow up. We derive the efficient influence function and show that it is uniformly bounded without requiring censoring survival probabilities to be uniformly bounded away from zero. A cross fitted one step estimator is asymptotically normal with consistently estimated variance under follow up weighted nuisance rate conditions. Simulations show little difference from overlap weighting when follow up is adequate, but substantially greater stability as follow up deteriorates. An analysis of the Rotterdam breast cancer cohort illustrates both the protection against weak follow up and the potential cost in precision.
\end{abstract}

\noindent\rule{16cm}{0.3pt}

\section{Introduction}\label{sec:intro}

Observational treatment comparisons require adequate overlap: patients in the
target population must have a meaningful chance of receiving either treatment.
When propensity scores approach zero or one, inverse probability weights become
extreme, estimators become unstable, and treatment effects for the full
population may cease to be regularly estimable
\citep{crump2009dealing,khan2010irregular}. Overlap weighting addresses this
problem by targeting patients for whom treatment assignment is most uncertain,
using the tilt $e(X)\{1-e(X)\}$ \citep{li2018balancing}. The approach has also
been extended to censored survival outcomes, including survival probabilities
and restricted mean survival times (RMSTs)
\citep{cheng2022overlap,zeng2023pseudo,cao2025rmst,xu2023weighted}.

With right-censored outcomes, treatment overlap is not the only support
required by the data. Estimation of an RMST contrast through a horizon $\tau$
also depends on the probability
\[
G_a(\tau\mid X)=\Pr(C>\tau\mid A=a,X)
\]
of remaining under observation. A patient may be at perfect treatment
equipoise, $e(X)=1/2$, yet have almost no chance of remaining uncensored through
$\tau$. This arises naturally, for example, with staggered entry and
administrative study closure. Ordinary overlap weights do not account for this
second source of limited support and may assign their largest weights where
follow-up information is weakest.

We show that this is a regularity problem, not merely finite-sample
instability. Pointwise positivity of treatment and censoring does not guarantee
a finite semiparametric efficiency bound for the overlap-weighted RMST effect.
Indeed, we construct a randomized-treatment model in which every censoring
probability is positive but the efficiency bound is infinite. A
bounded-support sequence shows the corresponding deterioration before this
limiting failure. We call the missing source of information
\emph{follow-up overlap}: whether outcomes through the restriction horizon can
realistically be observed for patients with a given covariate profile.

To account for both treatment and follow-up support, we propose
\emph{censoring-adjusted overlap} (CO) weighting. For fixed $c>0$, define
\[
h_c(X)
=
\left\{
\frac{\kappa_{1,c}(X)}{e(X)}
+
\frac{\kappa_{0,c}(X)}{1-e(X)}
\right\}^{-1},
\qquad
\kappa_{a,c}(X)
=
1+c\left\{\frac{1}{G_a(\tau\mid X)}-1\right\}.
\]
When $c=1$, the construction combines the probabilities of receiving treatment
and remaining under observation through $\tau$; as $c\downarrow0$, the target
approaches ordinary overlap weighting. Thus $c$ controls how strongly the
target retreats from covariate profiles with weak follow-up. Because the
weights depend only on the treatment and censoring mechanisms, the target is
defined at the design stage, before outcomes are analyzed.

The key inequality
\[
h_c(X)\le \frac{e_a(X)G_a(\tau\mid X)}{c}
\]
provides the cancellation that drives the theoretical results. We derive the
efficient influence function for the CO estimand and show that it is uniformly
bounded without requiring
$\inf_{a,x}G_a(\tau\mid x)>0$. The same structure removes inverse-censoring
amplification from the second-order remainder. Under stated nuisance-rate
conditions, a cross-fitted one-step estimator is asymptotically normal with
consistently estimated variance, with inference valid uniformly over a class
allowing follow-up probabilities to approach zero. Ordinary overlap weighting
does not have the same guarantee over this class.

CO changes the target population rather than attempting to recover the
ordinary overlap target through increasingly weak follow-up. This distinguishes
it from truncating inverse censoring weights, which modifies estimation of the
original target, and from recent approaches that select weighted estimands by
trading precision against target-population departure
\citep{barnard2026framework,orihara2026extrapolation}. We therefore make the
estimand change explicit by reporting retained target mass and target-population
distance and by examining a prespecified path of $c$ values.

Simulations show little difference between CO and ordinary overlap weighting
when follow-up is adequate. Under limited follow-up, overlap-weighted estimates
develop heavy tails, including under randomized treatment, whereas CO remains
stable with coverage near nominal. The Rotterdam breast cancer analysis also
shows that this protection need not improve precision: when patients with weak
follow-up carry substantial treatment information, moving away from them can
increase variance. Thus the contribution is regular inference under weak
follow-up support, not variance dominance.

The remainder of the paper develops the follow-up-overlap problem, introduces
the CO target and its diagnostics, establishes the corresponding regularity
and inference results, and evaluates the method in simulations and the
Rotterdam cohort. Proofs, numerical verification, and additional results are
provided in the Supplementary Materials.


\section{Follow-up overlap}\label{sec:problem}

\subsection{Setup and estimands}\label{sec:setup}

Let $X$ denote baseline covariates, $A\in\{0,1\}$ a binary treatment, $T^a$ the potential event time under treatment
$a$, $T=T^A$ the event time, and $C$ a right-censoring time. We observe $O=(X,A,\tilde T,\Delta)$, where
$\tilde T=\min(T,C)$ and $\Delta=1(T\le C)$. For a prespecified horizon $\tau>0$, the restricted mean survival time under
treatment $a$ for a patient with covariates $X$ is $\mu_a(X)=E\{\min(T^a,\tau)\mid X\}$. Write
$e_1(X)=e(X)=\mathrm{pr}(A=1\mid X)$ for the propensity score and $e_0(X)=1-e(X)$, and, for $a=0,1$,
\[
S_a(t\mid X)=\mathrm{pr}(T>t\mid A=a,X),\qquad
G_a(t\mid X)=\mathrm{pr}(C>t\mid A=a,X),
\]
with cumulative hazards $\Lambda_{T,a}$ and $\Lambda^C_a$. We make the following assumptions: consistency, $T=T^A$;
no unmeasured confounding, $T^a\perp\!\!\!\perp A\mid X$ for $a=0,1$; and conditionally independent censoring,
$T\perp\!\!\!\perp C\mid(A,X)$. Throughout, the censoring distribution is taken to be continuous; Appendix~A gives the
modifications needed when it has atoms.

\begin{condition}[Pointwise positivity]\label{cond:P}
$0<e(X)<1$ and $G_a(\tau\mid X)>0$ for $a=0,1$, almost surely.
\end{condition}

Condition~\ref{cond:P} imposes no lower bound on either probability. Under these assumptions,
$\mu_a(X)=\int_0^\tau S_a(t\mid X)\,dt$ is identified, and so is every weighted average treatment effect
\begin{equation}\label{eq:psih}
\psi_h=\frac{E\{h(X)\Delta(X)\}}{E\{h(X)\}},\qquad
\Delta(X)=\mu_1(X)-\mu_0(X),
\end{equation}
for a nonnegative tilting function $h$ with $E\{h(X)\}>0$. The choice $h\equiv1$ gives the average treatment effect, and
$h=e(1-e)$ gives the overlap-weighted effect \citep{li2018balancing}, which we denote $\psi_{\OW}$.

Two further conditions on the event-time distribution are used only where stated.

\begin{condition}[Event survival]\label{cond:E1}
$S_a(\tau\mid X)\ge s_0>0$ for $a=0,1$, almost surely.
\end{condition}

\begin{condition}[Non-degenerate residual survival]\label{cond:E2}
For some $s\in(0,\tau)$ and $\omega_{\min}>0$, almost surely for $a=0,1$ and all $u\le s$,
\[
\omega_a(u,X)=\mathrm{var}\{\min(T,\tau)\mid T\ge u,A=a,X\}\ge\omega_{\min}.
\]
\end{condition}

Condition~\ref{cond:E1} holds whenever the event hazard is bounded on $[0,\tau]$; it concerns the outcome, not the
censoring. Condition~\ref{cond:E2} rules out event-time distributions that become degenerate. It is used only to show that
a variance is infinite, in Theorem~\ref{thm:finite}(ii); neither the censoring-adjusted target nor any of its stability or
inference results require it.

\subsection{Efficient estimation for a fixed tilt}\label{sec:fixedtilt}

For a fixed tilt $h$, the efficient estimator of $\psi_h$ combines inverse probability of treatment weighting with the
augmented inverse probability of censoring weighted (AIPCW) transform of \citet{robins1992recovery}. Let
$Y_\tau=\min(T,\tau)$ denote the restricted outcome and $R_\tau=1(C\ge Y_\tau)$ its observation indicator. Both are
observable: $R_\tau=1$ exactly when $\Delta=1$ or $\tilde T\ge\tau$, and then $Y_\tau=U$, where
$U=\min(\tilde T,\tau)$. Let $N^C(u)=1(\tilde T\le u,\Delta=0)$. Within treatment arm $a$, let
\[
dM^C_a(u)
=
dN^C(u)-1(\tilde T\ge u)\,d\Lambda^C_a(u\mid X)
\]
denote the censoring martingale increment, and let
$m_a(u,X)=E\{\min(T,\tau)\mid T\ge u,A=a,X\}$. The transform is
\begin{equation}\label{eq:aipcw}
\phi_a(O)
=
\frac{R_\tau\,U}{G_a(U-\mid X)}
+
\int_0^{U}\frac{m_a(u,X)}{G_a(u-\mid X)}\,dM^C_a(u),
\end{equation}
which satisfies $E\{\phi_a(O)\mid A=a,X\}=\mu_a(X)$. The efficient influence function of $\psi_h$, treating $h$ as known,
is
\begin{equation}\label{eq:Dh}
\begin{aligned}
D_h(O)=\frac{1}{E\{h(X)\}}\Big[&
h(X)\{\Delta(X)-\psi_h\}\\
&+h(X)\Big\{
\frac{A}{e(X)}\{\phi_1(O)-\mu_1(X)\}
-\frac{1-A}{1-e(X)}\{\phi_0(O)-\mu_0(X)\}
\Big\}\Big],
\end{aligned}
\end{equation}
so any regular, asymptotically linear estimator of $\psi_h$ has asymptotic variance at least
\begin{equation}\label{eq:Vh}
V(h)=E\{D_h(O)^2\}
=
\frac{1}{E\{h(X)\}^2}\,
E\left[
h(X)^2\left\{
\frac{v_1(X)}{e(X)}
+
\frac{v_0(X)}{1-e(X)}
\right\}
+
h(X)^2\{\Delta(X)-\psi_h\}^2
\right],
\end{equation}
where $v_a(X)=\mathrm{var}\{\phi_a(O)\mid A=a,X\}$. When $h$ depends on the propensity score, as overlap weights do,
estimating $h$ adds to~\eqref{eq:Dh} the term
\[
\frac{
\{\Delta(X)-\psi_h\}\,
\partial_e h(X)\,
\{A-e(X)\}
}{
E\{h(X)\}
}.
\]
For $h=e(1-e)$ this term is bounded, because $|\partial_e h|\le1$ and $|\Delta-\psi_h|\le2\tau$, so it does not affect
whether $V(h)$ is finite.

The restricted mean survival time lies in $[0,\tau]$, so the outcome variance,
$\sigma_a^2(X)=\mathrm{var}\{\min(T^a,\tau)\mid X\}$, is at most $\tau^2/4$ and the effect-heterogeneity term
in~\eqref{eq:Vh} is at most $4\tau^2$. Once these bounded pieces are set aside, the remaining question is how censoring
inflates $v_a$.

\begin{lemma}[Censoring inflation]\label{lem:inflation}
Under the assumptions of Section~\ref{sec:setup},
\[
v_a(X)=\sigma_a^2(X)+C_a(X),\qquad
C_a(X)
=
\int_0^\tau
\omega_a(u,X)\,
\frac{S_a(u-\mid X)}{G_a(u-\mid X)}
\,d\Lambda^C_a(u\mid X).
\]
Moreover, for every $s\in(0,\tau)$,
\[
S_a(s\mid X)
\inf_{u\le s}\omega_a(u,X)
\left\{
\frac{1}{G_a(s\mid X)}-1
\right\}
\le
C_a(X)
\le
\frac{\tau^2}{4}
\left\{
\frac{1}{G_a(\tau\mid X)}-1
\right\}.
\]
\end{lemma}

The decomposition follows by writing
$\phi_a-\mu_a=(Y_\tau-\mu_a)-W_a$, where
\[
W_a
=
\int_0^{Y_\tau}
\{Y_\tau-m_a(u,X)\}
\frac{dM^C_a(u)}{G_a(u-\mid X)}
\]
is a censoring-martingale integral. Given $(T,A,X)$, both its integrand and its upper limit are fixed and $M^C_a$ is a
martingale, so $E(W_a\mid T,A,X)=0$; hence $W_a$ is uncorrelated with $Y_\tau$ given $(A,X)$
(Appendix~A.1). The bounds use the identity
$\int_0^t d\Lambda^C_a/G_a=1/G_a(t)-1$, valid for a continuous censoring law. The lemma says that censoring inflates the
variance contribution of a covariate profile by a factor that lies between the inverse follow-up probabilities at an
interior time $s$ and at the horizon $\tau$.

\subsection{When is the variance finite?}\label{sec:finite}

\begin{theorem}\label{thm:finite}
Assume Condition~\ref{cond:P} and $E\{h(X)\}>0$.
\begin{enumerate}
\item[(i)] If
\[
E\left\{
\frac{h(X)^2}{e(X)\,G_1(\tau\mid X)}
+
\frac{h(X)^2}{\{1-e(X)\}\,G_0(\tau\mid X)}
\right\}
<\infty,
\]
then $V(h)<\infty$.

\item[(ii)] Under Conditions~\ref{cond:E1} and~\ref{cond:E2}, if
\[
E\left\{
\frac{h(X)^2}{e_a(X)\,G_a(s\mid X)}
\right\}
=
\infty
\]
for some arm $a$ and the time $s$ of Condition~\ref{cond:E2}, then $V(h)=\infty$.
\end{enumerate}
\end{theorem}

For overlap weights, $h^2/e_1=e(1-e)^2$ and $h^2/e_0=e^2(1-e)$, so finiteness requires a moment of the inverse follow-up
probability weighted by the treatment-overlap factors. These factors control the propensity score, but they carry no
information about $G_a$.

\begin{corollary}[Overlap weighting under randomized treatment]\label{cor:rct}
Let $X\sim N(0,1)$, let treatment be randomized with $e(X)\equiv p\in(0,1)$, let the event hazard be constant, and let
the censoring hazard in at least one arm be $\lambda_C\exp(\gamma X)$ with $\lambda_C>0$ and $\gamma\ne0$. Then
Conditions~\ref{cond:P}--\ref{cond:E2} hold, yet $V(h_{\OW})=V(1)=\infty$.
\end{corollary}

\begin{proof}[Proof sketch]
With a constant propensity score, both tilts are constant. For every $s>0$,
\[
\frac{1}{G_a(s\mid X)}
=
\exp\{\lambda_C s\,e^{\gamma X}\},
\]
the exponential of a lognormal random variable, whose expectation is infinite. Theorem~\ref{thm:finite}(ii) applies.
Full details are in Appendix~A.
\end{proof}

Under this law, treatment overlap is as good as it can be, and every positivity condition holds pointwise, yet overlap
weighting has an infinite variance bound. The same divergence occurs under a logistic propensity score with a linear index
for every tilt $\{e(1-e)\}^{\nu}$, $\nu\ge0$ (Appendix~A). Treatment-overlap tilts that depend on $X$ only through
$e(X)$ therefore provide no general protection against poor follow-up overlap.

Because the observed-data model is locally nonparametric under conditionally independent censoring
\citep{gill1997coarsening}, the gradient~\eqref{eq:Dh} is the canonical gradient, and when it is not square-integrable the
semiparametric efficiency bound for $\psi_{\OW}$ is infinite.

The parameter remains identified; this is the censoring analogue of the irregular identification described by
\citet{khan2010irregular}, with the inverse probability of remaining uncensored in place of the inverse propensity score.

Bounded covariates do not remove the problem. Restricting $X$ to $|X|<b$ in Corollary~\ref{cor:rct} keeps
$V(h_{\OW})$ finite for each $b$, but it grows from 9.2 at $b=4$ to $8.5\times 10^{20}$ at $b=6$, with the same event-time
constants throughout (Appendix~A). This creates the potential for finite-sample behavior to be dominated by a small
number of patients whose probability of remaining under observation is extremely low; Section~\ref{sec:simulation} shows
this happening.


\section{The censoring-adjusted overlap target}\label{sec:target}

\subsection{Definition and interpretation}\label{sec:definition}

Overlap weights can be written as
\[
h_{\OW}(X)
=
\left\{
\frac{1}{e_1(X)}+\frac{1}{e_0(X)}
\right\}^{-1},
\]
a harmonic combination of the two treatment probabilities that downweights covariate profiles for which either treatment is unlikely. We extend this construction to account for the probability of remaining under observation through the analysis horizon.

\begin{definition}[Censoring-adjusted overlap target]\label{def:co}
For a fixed $c\in(0,1]$ and prespecified horizon $\tau$, define
\[
\kappa_{a,c}(X)
=
1+c\left\{
\frac{1}{G_a(\tau\mid X)}-1
\right\},
\qquad
h_c(X)
=
\left\{
\frac{\kappa_{1,c}(X)}{e(X)}
+
\frac{\kappa_{0,c}(X)}{1-e(X)}
\right\}^{-1}.
\]
The censoring-adjusted overlap (CO) estimand is
\[
\psi_c
=
\frac{E\{h_c(X)\Delta(X)\}}
     {E\{h_c(X)\}}.
\]
\end{definition}

The target population has covariate density proportional to $h_c(X)$ times the study-population density. Importantly, $h_c$ depends only on the treatment and censoring mechanisms, not on the event-time distribution, so the target population can be characterized before outcomes are analyzed.

The construction also has a direct overlap interpretation. Define
\[
\widetilde G_{a,c}(X)
=
\frac{G_a(\tau\mid X)}
     {c+(1-c)G_a(\tau\mid X)}.
\]
Then
\[
\frac{e_a(X)}{\kappa_{a,c}(X)}
=
e_a(X)\widetilde G_{a,c}(X),
\qquad
G_a(\tau\mid X)
\le
\widetilde G_{a,c}(X)
\le1,
\]
and
\[
h_c(X)
=
\left\{
\frac{1}{e_1(X)\widetilde G_{1,c}(X)}
+
\frac{1}{e_0(X)\widetilde G_{0,c}(X)}
\right\}^{-1}.
\]
When $c=1$, $\widetilde G_{a,1}(X)=G_a(\tau\mid X)$ and
\[
e_a(X)G_a(\tau\mid X)
=
\mathrm{pr}(A=a,C>\tau\mid X).
\]
Thus $c=1$ defines overlap in the joint probabilities of receiving each treatment and remaining under observation through $\tau$. As $c$ decreases, the follow-up adjustment is softened and $h_c$ approaches ordinary overlap weighting.

The harmonic construction differs from simply multiplying overlap weights by $G_1(\tau\mid X)G_0(\tau\mid X)$. The latter downweights a profile whenever follow-up is weak in either arm, regardless of the corresponding treatment probability. CO instead combines treatment and follow-up information arm by arm. We compare these constructions empirically in Section~\ref{sec:simulation}.

\subsection{Basic properties}\label{sec:properties}

\begin{proposition}\label{prop:properties}
Let $c\in(0,1]$ and suppose Condition~\ref{cond:P} holds. Then, almost surely:
\begin{enumerate}
\item[(i)] if $G_1(\tau\mid X)=G_0(\tau\mid X)=1$, then
\[
h_c(X)=e(X)\{1-e(X)\},
\]
and $h_c(X)\to e(X)\{1-e(X)\}$ as $c\downarrow0$;

\item[(ii)] $h_c(X)$ is nonincreasing in $c$;

\item[(iii)]
\[
0<h_c(X)
\le
\min\{e(X),1-e(X)\},
\qquad
h_c(X)
\le
\frac{e_a(X)G_a(\tau\mid X)}{c},
\quad a=0,1;
\]

\item[(iv)]
\[
v_a(X)
\le
\frac{\tau^2}{4G_a(\tau\mid X)},
\qquad
\kappa_{a,c}(X)
\ge
\frac{c}{G_a(\tau\mid X)},
\]
and hence
\[
\frac{v_a(X)}{\kappa_{a,c}(X)}
\le
\frac{\tau^2}{4c}.
\]
\end{enumerate}
\end{proposition}

Parts (i)--(iii) follow from
$\kappa_{a,c}=1-c+c/G_a(\tau\mid X)$ and
$h_c\le e_a/\kappa_{a,c}$. Part (iv) follows from
Lemma~\ref{lem:inflation} and $\sigma_a^2(X)\le\tau^2/4$.

Proposition~\ref{prop:properties}(iii) is central to the construction: as follow-up in arm $a$ deteriorates, the CO tilt decreases at least proportionally to $G_a(\tau\mid X)$. Part (iv) shows the corresponding cancellation of the censoring inflation identified in Lemma~\ref{lem:inflation}. These properties require neither a uniform lower bound on $G_a(\tau\mid X)$ nor Condition~\ref{cond:E2}.

When follow-up is nearly complete, $h_c$ approaches ordinary overlap weighting. Under limited follow-up, the two tilts generally define different populations. Consequently, $\psi_c$ and $\psi_{\OW}$ differ only through treatment-effect heterogeneity across covariate profiles that are weighted differently by the two targets. In particular, they coincide when $\Delta(X)$ is constant.

\subsection{The protection path}\label{sec:cpath}

The constant $c$ controls the strength of the follow-up adjustment. Our theoretical results hold for any fixed
$c\in[c_{\min},1]$, $c_{\min}>0$, with bounds that scale as $1/c$. Thus $c$ is part of the estimand specification rather than a data-adaptive tuning parameter; allowing it to approach zero with sample size would remove the protection established below.

We use $c=0.1$ as the primary specification and $c\in\{0.3,1\}$ as prespecified sensitivity values. In the simulation settings of Section~\ref{sec:simulation}, $c=0.1$ remained close to ordinary overlap weighting when follow-up was adequate while providing substantial protection when follow-up deteriorated. The larger values provide progressively stronger adjustment.

The path $c\mapsto\psi_c$ can be reported together with the relative target mass
\[
\frac{E\{h_c(X)\}}{E\{h_{\OW}(X)\}}
\]
and the total-variation distance between the corresponding target populations. These quantities describe how strongly accounting for follow-up changes the population being targeted. Because $G_a(\tau\mid X)$ depends on $\tau$, the CO target is horizon-specific. If a common target is desired across several horizons, the tilt may instead be fixed using the largest horizon considered.

\subsection{Scope and design diagnostics}\label{sec:scope}

CO addresses \emph{limited} follow-up overlap: $G_a(\tau\mid X)$ remains positive almost surely but may approach zero. It does not resolve structural absence of follow-up, where $G_a(\tau\mid X)=0$ on a set of positive probability. Such horizons fall outside Condition~\ref{cond:P} and the results developed here. In the application of Section~\ref{sec:application}, we therefore excluded a horizon with structural loss of follow-up before examining the outcome analysis.

Because the CO tilt depends only on the treatment and censoring mechanisms, follow-up overlap can be assessed at the design stage. Useful diagnostics include the joint distribution of $\hat e(X)$ and $\hat G_a(\tau\mid X)$, particularly among patients near treatment equipoise, together with the relative target mass, total-variation distance from the OW population, and covariate composition of the CO target along the prespecified $c$ path. These diagnostics characterize how much the target population changes when follow-up information is incorporated.


\section{Regularity of the CO target}\label{sec:regularity}

\subsection{Oracle stability}\label{sec:oracle}

We first treat the tilt as known. The cancellation in
Proposition~\ref{prop:properties}(iv) is sufficient to control the
efficiency bound.

\begin{theorem}[Oracle stability]\label{thm:oracle}
Assume Condition~\ref{cond:P} and fix $c\in(0,1]$. Then
$E\{h_c(X)\}>0$ and
\[
V(h_c)
\le
\frac{\tau^2\{1/(4c)+4\}}{E\{h_c(X)\}}.
\]
\end{theorem}

\begin{proof}
By Proposition~\ref{prop:properties}(iv),
$v_a\le\{\tau^2/(4c)\}\kappa_{a,c}$, and therefore
\[
h_c^2
\left(
\frac{v_1}{e}
+
\frac{v_0}{1-e}
\right)
\le
\frac{\tau^2}{4c}
h_c^2
\left(
\frac{\kappa_{1,c}}{e}
+
\frac{\kappa_{0,c}}{1-e}
\right)
=
\frac{\tau^2}{4c}h_c.
\]
The heterogeneity term in~\eqref{eq:Vh} is at most
$4\tau^2h_c^2\le4\tau^2h_c$. Taking expectations and dividing by
$E(h_c)^2$ gives the result. Condition~\ref{cond:P} implies
$0<h_c<\infty$ almost surely, and hence $E(h_c)>0$.
\end{proof}

No moment condition on $1/G_a(\tau\mid X)$ appears in this bound. In
particular, over a class of distributions satisfying
$E\{h_c(X)\}\ge m_0>0$, the oracle variance is bounded by
$\tau^2\{1/(4c)+4\}/m_0$. The analogous mass condition for ordinary
overlap weighting does not control its variance, as shown by
Corollary~\ref{cor:rct}.

\subsection{Efficient influence function}\label{sec:eif}

The CO estimand is a moving target because $h_c$ depends on the
propensity score and censoring distribution. Its influence function
therefore contains contributions from perturbing both mechanisms.

\begin{lemma}[Censoring gradient]\label{lem:cgrad}
Fix $a$ and $x$, and let
\[
y_a(u\mid x)
=
S_a(u-\mid x)G_a(u-\mid x)
\]
denote the probability of being at risk at time $u$ in arm $a$.
Under a continuous censoring law, the canonical gradient of
$G_a(\tau\mid x)$ in the conditional model given $(A=a,X=x)$ is
\[
\gamma_a(O)
=
-G_a(\tau\mid x)
\int_0^\tau
\frac{dM_a^C(u)}{y_a(u\mid x)}.
\]
\end{lemma}

The proof is given in Appendix~B. Write
\[
H(e,g_1,g_0)
=
\left\{
\frac{\kappa_c(g_1)}{e}
+
\frac{\kappa_c(g_0)}{1-e}
\right\}^{-1},
\qquad
\kappa_c(g)=1-c+\frac{c}{g},
\]
so that
$h_c(X)=H\{e(X),G_1(\tau\mid X),G_0(\tau\mid X)\}$. The derivatives
needed below are
\begin{equation}\label{eq:Hderiv}
H_e
=
h_c^2
\left\{
\frac{\kappa_{1,c}}{e^2}
-
\frac{\kappa_{0,c}}{(1-e)^2}
\right\},
\qquad
H_{g_a}
=
\frac{c\,h_c^2}
     {e_aG_a(\tau\mid X)^2}.
\end{equation}

\begin{theorem}[Efficient influence function]\label{thm:eif}
Assume Condition~\ref{cond:P} and a continuous censoring law. The
efficient influence function of $\psi_c$ under the observed-data model
of Section~\ref{sec:setup} is
\begin{equation}\label{eq:Dco}
\begin{aligned}
D_{\CO}(O)
=
\frac{1}{E\{h_c(X)\}}
\Bigg[&
h_c(X)\{\Delta(X)-\psi_c\}\\
&+
h_c(X)
\left\{
\frac{A}{e(X)}\{\phi_1(O)-\mu_1(X)\}
-
\frac{1-A}{1-e(X)}\{\phi_0(O)-\mu_0(X)\}
\right\}\\
&+
\{\Delta(X)-\psi_c\}
H_e(X)\{A-e(X)\}\\
&-
c\{\Delta(X)-\psi_c\}h_c(X)^2
\sum_{a=0,1}
\frac{1(A=a)}
     {e_a(X)^2G_a(\tau\mid X)}
\int_0^\tau
\frac{dM_a^C(u)}
     {y_a(u\mid X)}
\Bigg].
\end{aligned}
\end{equation}
\end{theorem}

The four lines of~\eqref{eq:Dco} arise from perturbations of the
covariate distribution, event-time distribution, treatment mechanism,
and censoring mechanism, respectively. Appendix~B derives the
gradient along a regular parametric submodel; numerical checks along
each score direction are reported in Appendix~C.

The outcome component can equivalently be written as the
event-martingale integral
\[
\phi_a(O)-\mu_a(X)
=
-
\int_0^\tau
\frac{\int_u^\tau S_a(t\mid X)\,dt}
     {S_a(u\mid X)G_a(u-\mid X)}
\,dM_a^T(u),
\]
where
\[
dM_a^T(u)
=
dN^T(u)
-
1(\tilde T\ge u)\,d\Lambda_{T,a}(u\mid X),
\qquad
N^T(u)=1(\tilde T\le u,\Delta=1).
\]
The event and censoring martingales are orthogonal, yielding the
following decomposition.

\begin{corollary}[Variance decomposition]\label{cor:decomp}
Under the conditions of Theorem~\ref{thm:eif}, the four components of
\eqref{eq:Dco} are mutually uncorrelated and
\[
\begin{aligned}
E\{h_c(X)\}^2E(D_{\CO}^2)
={}&
E\big[h_c^2(\Delta-\psi_c)^2\big]\\
&+
E\big[(\Delta-\psi_c)^2H_e^2e(1-e)\big]\\
&+
E\left[
h_c^2
\left(
\frac{v_1}{e}
+
\frac{v_0}{1-e}
\right)
\right]\\
&+
c^2E\left[
(\Delta-\psi_c)^2h_c^4
\sum_{a=0,1}
\frac{1}
     {e_a^3G_a(\tau\mid X)^2}
\int_0^\tau
\frac{d\Lambda_a^C(u\mid X)}
     {y_a(u\mid X)}
\right].
\end{aligned}
\]
\end{corollary}

The decomposition separates variation due to the target population,
treatment mechanism, censored event outcomes, and estimation of the
censoring-dependent target.

\subsection{Bounded influence function}\label{sec:bounded}

The CO construction controls not only the oracle variance but the
influence function of the moving target.

\begin{proposition}[Bounded influence function]\label{prop:bounded}
Assume Conditions~\ref{cond:P} and~\ref{cond:E1}, a continuous
censoring law, and $c\in(0,1]$. Then
\[
\sup_o |D_{\CO}(o)|
\le
\frac{\tau}{E\{h_c(X)\}}
\left(
5+\frac{1}{c}+\frac{2}{cs_0}
\right).
\]
\end{proposition}

\begin{proof}[Proof sketch]
The AIPCW representation satisfies
\[
|\phi_a(O)|
\le
\frac{\tau}{G_a(U\mid X)}.
\]
Together with Proposition~\ref{prop:properties}(iii),
\[
\frac{h_c}{e_a}
\le
\frac{G_a(\tau\mid X)}{c}
\le
\frac{G_a(U\mid X)}{c},
\]
which removes the inverse follow-up probability from the outcome
component. Similarly, the censoring-gradient contribution is bounded
using $S_a(\tau\mid X)\ge s_0$ and
$H_{g_a}/e_a\le1/c$. Finally,
$|\Delta-\psi_c|\le2\tau$, $h_c\le1$, and $|H_e|\le1$ control the
remaining components. Combining these bounds gives the stated
constant. Details are in Appendix~B; the derivative and numerical
bound checks are reported in Appendix~C.
\end{proof}

Unlike ordinary overlap weighting, Proposition~\ref{prop:bounded}
requires no uniform lower bound on $G_a(\tau\mid X)$. For OW,
$h_{\OW}/e_a$ does not vanish with $G_a$, leaving the inverse
follow-up factor in the outcome contribution; Corollary~\ref{cor:rct}
shows that the resulting influence function need not be
square-integrable. For CO, the bound depends only on
$\tau$, $c$, $s_0$, and the target mass $E\{h_c(X)\}$. This boundedness
is the key regularity property used for estimation and inference in
Section~\ref{sec:inference}.


\section{Estimation and inference}\label{sec:inference}

\subsection{Estimation}\label{sec:estimator}

Estimation requires the propensity score $e(X)$ and, in each treatment
arm, the event and censoring survival functions $S_a(t\mid X)$ and
$G_a(t\mid X)$. We use $K$-fold cross-fitting
\citep{chernozhukov2018dml}, allowing these nuisance functions to be
estimated flexibly without imposing Donsker conditions.

Let $I_1,\ldots,I_K$ denote folds of approximately equal size. For
each fold $I_k$, estimate $(e,S_a,G_a)$ using observations outside
$I_k$ and evaluate the fitted nuisance functions on observations in
$I_k$. We truncate $\hat e(X)$ to $[\epsilon_e/2,1-\epsilon_e/2]$, with $\epsilon_e$ as in Condition~4 below, and $\hat S_a(\tau\mid X)$ below at
$s_0/2$. For each observation, let $\hat h_i$, $\hat H_{e,i}$ and
$\hat H_{g_a,i}$ denote the estimated tilt and its derivatives,
$\hat\phi_{a,i}$ the AIPCW transform in~\eqref{eq:aipcw},
$\hat\Delta_i=\hat\mu_1(X_i)-\hat\mu_0(X_i)$, and
$\hat\gamma_{a,i}$ the censoring gradient of
Lemma~\ref{lem:cgrad}.

Define
\[
\tilde\psi_c
=
\frac{
\sum_{i=1}^n
\hat h_i\{\hat\phi_{1,i}-\hat\phi_{0,i}\}
}{
\sum_{i=1}^n\hat h_i
}.
\]
This is the AIPCW estimator obtained by treating the estimated tilt as
fixed. The one-step estimator additionally accounts for estimation of
the target population:
\begin{equation}\label{eq:onestep}
\hat\psi_c
=
\tilde\psi_c
+
\frac{
\displaystyle
\sum_{i=1}^n
(\hat\Delta_i-\tilde\psi_c)
\left\{
\hat H_{e,i}(A_i-\hat e_i)
+
\sum_{a=0,1}
\frac{
\hat H_{g_a,i}1(A_i=a)\hat\gamma_{a,i}
}{
\hat e_{a,i}
}
\right\}
}{
\displaystyle
\sum_{i=1}^n\hat h_i
}.
\end{equation}

Let $\hat D_i$ denote the influence function~\eqref{eq:Dco} evaluated
at the cross-fitted nuisance estimates, with $E\{h_c(X)\}$ replaced
by $n^{-1}\sum_i\hat h_i$. We estimate the asymptotic variance by
\[
\hat V
=
\frac1n
\sum_{i=1}^n
(\hat D_i-\bar D)^2,
\qquad
\bar D=\frac1n\sum_{i=1}^n\hat D_i,
\]
giving the Wald interval
\[
\hat\psi_c
\pm
1.96\sqrt{\frac{\hat V}{n}}.
\]

\subsection{Second-order behavior}\label{sec:remainder}

The main difficulty in estimation under weak follow-up is that
ordinary AIPCW remainders contain inverse censoring probabilities.
The CO tilt removes this amplification.

Write $P$ for the true distribution and $\hat P$ for a distribution
with nuisance functions $(\hat e,\hat S_a,\hat G_a)$ and the same
covariate distribution as $P$. The one-step expansion is
\begin{equation}\label{eq:expansion}
\hat\psi_c-\psi_c
=
(\mathbb P_n-P)D_{\CO,P}
+
(\mathbb P_n-P)
(D_{\CO,\hat P}-D_{\CO,P})
+
R_2(\hat P,P),
\end{equation}
where
\[
R_2(\hat P,P)
=
\psi_c(\hat P)-\psi_c(P)
+
P D_{\CO,\hat P}.
\]

The first source of second-order error is the mixed bias of the
AIPCW transform.

\begin{lemma}[Mixed bias]\label{lem:mixed}
Suppose the event and censoring distributions are continuous. Then
\[
r_a(X)
=
E_P\{\hat\phi_a(O)\mid A=a,X\}-\mu_a(X)
\]
satisfies
\[
\begin{aligned}
r_a(X)
=
\int_0^\tau&
\left\{
\int_u^\tau
\hat S_a(t\mid X)\,dt
\right\}
\frac{S_a(u\mid X)}
     {\hat S_a(u\mid X)}
\frac{\hat G_a(u\mid X)-G_a(u\mid X)}
     {\hat G_a(u\mid X)}
\\
&\hspace{25mm}\times
d\{\Lambda_{T,a}-\hat\Lambda_{T,a}\}(u\mid X).
\end{aligned}
\]
\end{lemma}

The mixed bias is a product of censoring-model and event-hazard
errors, but its usual representation contains
$1/\hat G_a(u\mid X)$. For ordinary overlap weighting this inverse
follow-up factor remains. For CO, Proposition~\ref{prop:properties}(iii)
gives, for every $u\le\tau$,
\begin{equation}\label{eq:cancel}
\frac{
\hat h(X)
}{
\hat e_a(X)\hat G_a(u\mid X)
}
\le
\frac{
\hat G_a(\tau\mid X)
}{
c\,\hat G_a(u\mid X)
}
\le
\frac1c.
\end{equation}
Thus the CO remainder involves absolute censoring error rather than
error amplified by an inverse follow-up probability. The same
cancellation controls the terms arising from estimation of the tilt;
details are given in Appendix~B.

To state the remainder bound, let
\[
\delta_a(X)
=
\log\hat G_a(\tau\mid X)
-
\log G_a(\tau\mid X),
\]
and, for fixed $M>0$, define
\[
\begin{aligned}
\rho_1
={}&
\max_a
E\left[
\int_0^\tau
|\hat G_a-G_a|(u\mid X)
\,
\big|
d(\hat\Lambda_{T,a}-\Lambda_{T,a})(u\mid X)
\big|
\right]
\\
&+
\max_a
E\big(
|\hat e-e|\,|\hat\mu_a-\mu_a|
\big),
\\[1mm]
\rho_2
={}&
E\{(\hat e-e)^2\},
\\[1mm]
\rho_3
={}&
\max_a
E\left\{
G_a(\tau\mid X)\delta_a(X)^2
\right\}
\\
&+
\max_a
E\left[
\int_0^\tau
G_a(u\mid X)
\left\{
|\hat S_a-S_a|(u\mid X)
+
|\hat e-e|
\right\}
\,
\big|
d(\hat\Lambda_a^C-\Lambda_a^C)(u\mid X)
\big|
\right],
\\[1mm]
\rho_4
={}&
\max_a
\Pr_X\{|\delta_a(X)|>M\},
\\[1mm]
\rho_5
={}&
E\Bigg[
\Big\{
|\hat\Delta-\Delta|
+
|\psi_c(\hat P)-\psi_c|
\Big\}
\\
&\hspace{13mm}\times
\left\{
|\hat e-e|
+
\sum_{a=0,1}
\int_0^\tau
G_a(u\mid X)
\big|
d(\hat\Lambda_a^C-\Lambda_a^C)(u\mid X)
\big|
\right\}
\Bigg].
\end{aligned}
\]

\begin{condition}[Treatment positivity]\label{cond:trt}
For some $\epsilon_e>0$,
\[
\epsilon_e<e(X)<1-\epsilon_e
\]
almost surely.
\end{condition}

\begin{proposition}[Remainder bound]\label{prop:remainder}
Assume Conditions~\ref{cond:P}, \ref{cond:E1} and~\ref{cond:trt},
fix $c\in(0,1]$, and suppose
$E\{h_c(X)\}\ge m_0>0$. For nuisance estimators truncated as in
Section~\ref{sec:estimator},
\[
|R_2(\hat P,P)|
\le
C\sum_{j=1}^5\rho_j,
\]
where $C$ depends only on
$(\tau,c,s_0,\epsilon_e,m_0,M)$. No uniform lower bound on
$G_a$ or $\hat G_a$ is required.
\end{proposition}

The quantities $\rho_1$ and $\rho_5$ contain products of nuisance
errors, while $\rho_2$ and $\rho_3$ contain quadratic and product
errors in the mechanisms that define the target. Importantly, censoring-model errors
in $\rho_3$ are weighted by $G_a$: the required accuracy weakens in
regions where follow-up itself becomes scarce. The term $\rho_4$
controls regions on which the estimated and true censoring
probabilities differ by more than a fixed multiplicative factor.
The exact remainder decomposition and derivative bounds are given in
Appendix~B.

\subsection{Asymptotic inference}\label{sec:theory}

The rate conditions above are supplemented by a rate-free consistency
condition.

\begin{condition}[Nuisance consistency]\label{cond:V}
For every $\eta>0$,
\[
\Pr_X\{|\hat e(X)-e(X)|>\eta\}\to_p0,
\]
\[
\Pr_X\left\{
\max_a\sup_{t\le\tau}
|\hat S_a(t\mid X)-S_a(t\mid X)|
>\eta
\right\}
\to_p0,
\]
and
\[
\Pr_X\left\{
\max_a\sup_{t\le\tau}
|
\log\hat G_a(t\mid X)
-
\log G_a(t\mid X)
|
>\eta
\right\}
\to_p0.
\]
Here $\Pr_X$ is taken over a new covariate value with the fitted
nuisance functions held fixed.
\end{condition}

The supremum in Condition~\ref{cond:V} is over time for each
covariate value; convergence is in probability over $X$. In
particular, the condition does not require uniform convergence over
the covariate space.

\begin{condition}[Rates]\label{cond:rates}
For some fixed $M>0$,
\[
\rho_j=o_p(n^{-1/2}),
\qquad j=1,\ldots,5.
\]
\end{condition}

\begin{theorem}[Asymptotic inference]\label{thm:an}
Assume Conditions~\ref{cond:P}, \ref{cond:E1},
\ref{cond:trt}, \ref{cond:V} and~\ref{cond:rates}, a continuous
censoring law, fixed $c\in(0,1]$, and
$E\{h_c(X)\}>0$. Then
\[
\hat\psi_c-\psi_c
=
\frac1n
\sum_{i=1}^n
D_{\CO}(O_i)
+
o_p(n^{-1/2}),
\]
and
\[
\sqrt n(\hat\psi_c-\psi_c)
\rightsquigarrow
N\{0,V_c\},
\qquad
V_c=E\{D_{\CO}(O)^2\}.
\]
Moreover,
\[
\hat V\to_p V_c.
\]
\end{theorem}

\begin{proof}[Proof sketch]
Proposition~\ref{prop:remainder} gives
$R_2(\hat P,P)=o_p(n^{-1/2})$. Conditional on the training folds,
cross-fitting and Chebyshev's inequality yield
\[
(\mathbb P_n-P)
(D_{\CO,\hat P}-D_{\CO,P})
=
O_p\left(
\frac{
\|D_{\CO,\hat P}-D_{\CO,P}\|_2
}{
\sqrt n
}
\right).
\]
Condition~\ref{cond:V} gives pointwise convergence in probability of
the estimated influence function at a new observation.
Proposition~\ref{prop:bounded}, applied also to the truncated
estimated nuisances, supplies a common bound; hence bounded
convergence gives
\[
\|D_{\CO,\hat P}-D_{\CO,P}\|_2=o_p(1).
\]
The asymptotic linear expansion and central limit theorem follow.
The same boundedness, together with Cauchy--Schwarz, yields
$\hat V\to_p V_c$. Full details are in Appendix~B.
\end{proof}

Theorem~\ref{thm:an} requires neither a Donsker condition nor a
uniform lower bound on the censoring survival probabilities. The
former is avoided by cross-fitting; the latter follows from the
bounded influence function and the cancellation in
\eqref{eq:cancel}. In contrast, the corresponding OW influence
function need not have a finite second moment
(Corollary~\ref{cor:rct}).

The boundedness result also yields uniform inference. Let
\[
\mathcal P
=
\mathcal P(
\tau,c_{\min},s_{\min},
\epsilon_e,m_0,v_{\min}
)
\]
denote a class of laws satisfying Condition~\ref{cond:P},
\[
S_a(\tau\mid X)\ge s_{\min},\qquad
\epsilon_e<e(X)<1-\epsilon_e,
\]
\[
E_P\{h_c(X)\}\ge m_0,
\qquad
\operatorname{var}_P(D_{\CO})\ge v_{\min},
\]
with $c\in[c_{\min},1]$.

\begin{theorem}[Uniform validity]\label{thm:uniform}
Over $\mathcal P$,
\[
\sup_{P\in\mathcal P}
\sup_o
|D_{\CO,P}(o)|
\le
K
=
\frac{\tau}{m_0}
\left(
5+\frac1{c_{\min}}
+\frac{2}{c_{\min}s_{\min}}
\right).
\]
If Conditions~\ref{cond:V} and~\ref{cond:rates} hold uniformly over
$\mathcal P$, then
\[
\sup_{P\in\mathcal P}
\sup_{z\in\mathbb R}
\left|
\Pr_P
\left\{
\frac{
\sqrt n(\hat\psi_c-\psi_c)
}{
\hat V^{1/2}
}
\le z
\right\}
-
\Phi(z)
\right|
\longrightarrow0.
\]
\end{theorem}

The uniform influence-function bound follows directly from
Proposition~\ref{prop:bounded}. Together with
$\operatorname{var}_P(D_{\CO})\ge v_{\min}$, it gives a uniform
Berry--Esseen bound, while the uniform nuisance conditions control
the second-order and variance-estimation terms. The contrast with OW
is sharp: the bounded-support sequence following
Corollary~\ref{cor:rct} satisfies fixed treatment-overlap and
event-survival constants while its OW efficiency bound diverges.

\subsection{Practical scope}\label{sec:limits}

The preceding results protect against small follow-up probabilities,
not against persistent failure to estimate them. Condition~\ref{cond:V}
excludes censoring models that systematically overstate follow-up on
a covariate region of nonvanishing probability. In such a region,
the estimated CO tilt may fail to provide the intended downweighting.
The sensitivity of CO to this form of misspecification is examined in Appendix~D.

Truncating small estimated censoring probabilities addresses the same
instability differently. Replacing $\hat G_a$ by
$\max(\hat G_a,g_{\min})$ modifies the estimator while retaining the
nominal target and can therefore introduce bias for that target. CO
instead changes the target population explicitly. Section~\ref{sec:simulation}
compares these approaches relative to their respective estimands.

The theoretical results above are stated for the cross-fitted
estimator, permitting flexible nuisance estimation without
empirical-process restrictions. In the simulations and application,
the nuisance functions are estimated using parametric and Cox models,
which belong to Donsker classes; these models are therefore fitted on
the full sample.


\section{Simulation study}\label{sec:simulation}

\subsection{Design}\label{sec:simdesign}

We designed the simulations to separate limited treatment overlap from
limited follow-up overlap. Covariates $X=(X_1,X_2,X_3)$ were independent
standard normal variables truncated to $|X_j|<3.5$. Treatment followed
\[
\operatorname{logit}\{e(X)\}=\alpha X_1,
\]
with $\alpha=0$ for randomized treatment and $\alpha=2.5$ for limited
treatment overlap. Event times were exponential with hazard
\[
\lambda_a(X)
=
0.2\exp\left\{
0.5\tanh X_1+
0.5\tanh X_2+
a(-0.3+0.5\tanh X_2)
\right\},
\]
so that the treatment effect varied with $X_2$.

Censoring times were exponential with hazard
\[
\lambda_{C0}\exp(\gamma X_2+0.3a),
\]
where $\gamma=0.3$ represented good follow-up overlap and $\gamma=1.4$
limited follow-up overlap. For each design, $\lambda_{C0}$ was chosen
to give 35\% marginal censoring before $\tau=3$ and was then held fixed
across horizons. Consequently, the two follow-up regimes have the same
marginal censoring proportion at $\tau=3$ but differ in how censoring
is distributed across covariate profiles. Because $X_2$ also modifies
the treatment effect, the design makes any change in estimand induced
by retreating from poorly followed covariate profiles visible.

We crossed the two treatment-overlap regimes, two follow-up regimes,
and horizons $\tau\in\{1,3,5\}$. For each of the 12 configurations,
we generated 2000 samples of size $n=2000$. The design, estimators,
and number of replicates were fixed before the final run. The
propensity score was estimated by logistic regression, and
arm-specific event and censoring hazards by exponential regression
using the correctly specified covariate functions.

We compared inverse probability weighting ($h\equiv1$), overlap
weighting (OW), CO weighting with $c\in\{0.1,0.3,1\}$, the product
tilt
\[
e(X)\{1-e(X)\}G_1(\tau\mid X)G_0(\tau\mid X),
\]
and OW with the estimated censoring survival function truncated below
at fixed values (0.05 or 0.10) or at its first or fifth percentile.
All methods used one-step estimation with influence-function standard
errors and Wald intervals. True values of the corresponding estimands
were approximated using $10^6$ covariate draws.

Each estimator was evaluated against its own estimand; truncated OW
estimators were evaluated against the OW estimand they were intended
to estimate. We report bias, empirical standard deviation, coverage,
and kurtosis. To characterize tail behavior more robustly, we also
report
\[
\mathrm{TI}
=
\frac{q_{0.99}(|\hat\psi-\psi|)}
     {q_{0.50}(|\hat\psi-\psi|)},
\]
which equals approximately 3.82 under normality. Relative precision is
summarized by the empirical variance ratio (VR) to OW, and target
movement by the total-variation (TV) distance from the OW target
population.

\subsection{Results}\label{sec:simresults}

Table~\ref{tab:sim} compares OW with the primary CO analysis
($c=0.1$) in all 12 configurations. Results for the remaining values
of $c$, the product tilt, IPW, and truncation estimators are reported
in Appendix~D.

\begin{sidewaystable}[!htbp]
\caption{simulation study comparing overlap weighting (OW) and
censoring-adjusted overlap weighting with $c=0.1$ (CO), with $n=2000$
and 2000 replicates per cell. $G_{1,0.05}$ is the fifth percentile of
$G_1(\tau\mid X)$. Bias and SD are multiplied by 100 on the restricted
mean survival time scale and are computed relative to each estimator's
own target. Cover is coverage (\%) of the 95\% influence-function
interval. Kurt is kurtosis of the estimates. TI is
$q_{0.99}/q_{0.50}$ of the absolute error (3.82 under normality).
VR is the variance of CO relative to OW, with a 95\% Monte Carlo
interval from 500 resamples of the replicates. TV is the
total-variation distance between the CO and OW target populations.}
\label{tab:sim}

\centering
\footnotesize
\begin{tabular}{lllrrrrrrrrrrrlr}
\toprule
 & \multicolumn{2}{c}{Overlap} & &
 \multicolumn{5}{c}{OW} &
 \multicolumn{5}{c}{CO ($c=0.1$)} & & \\
\cmidrule(lr){2-3}\cmidrule(lr){5-9}\cmidrule(lr){10-14}
$\tau$ & Treatment & Follow-up & $G_{1,0.05}$
& Bias & SD & Cover & Kurt & TI
& Bias & SD & Cover & Kurt & TI & VR & TV\\
\midrule
1 & Good & Good
& 0.767 & 0.0 & 1.0 & 95.2 & 3.1 & 4.03
& 0.0 & 1.0 & 95.2 & 3.1 & 4.02
& 1.00 [1.00, 1.00] & 0.00 \\
& Good & Limited
& 0.316 & 0.0 & 1.1 & 95.2 & 3.9 & 3.95
& 0.0 & 1.1 & 95.2 & 3.2 & 4.01
& 0.89 [0.85, 0.93] & 0.02 \\
& Limited & Good
& 0.768 & 0.0 & 1.5 & 95.2 & 3.0 & 3.87
& 0.0 & 1.5 & 95.1 & 3.0 & 3.89
& 1.00 [1.00, 1.00] & 0.00 \\
& Limited & Limited
& 0.316 & 0.0 & 1.5 & 94.5 & 3.4 & 4.00
& 0.0 & 1.5 & 94.6 & 3.0 & 3.88
& 0.92 [0.88, 0.95] & 0.02 \\
\addlinespace

3 & Good & Good
& 0.452 & 0.1 & 4.6 & 94.2 & 3.2 & 3.85
& 0.1 & 4.6 & 94.2 & 3.2 & 3.85
& 0.99 [0.99, 0.99] & 0.01 \\
& Good & Limited
& 0.032 & 0.3 & 5.9 & 94.8 & 7.9 & 4.83
& 0.2 & 4.6 & 95.3 & 3.1 & 3.89
& 0.60 [0.53, 0.66] & 0.08 \\
& Limited & Good
& 0.452 & 0.2 & 6.3 & 95.6 & 3.0 & 3.79
& 0.2 & 6.2 & 95.5 & 3.0 & 3.79
& 0.99 [0.99, 1.00] & 0.01 \\
& Limited & Limited
& 0.032 & 0.0 & 8.9 & 95.0 & 20.2 & 4.56
& 0.1 & 6.5 & 94.4 & 3.0 & 3.80
& 0.53 [0.44, 0.64] & 0.08 \\
\addlinespace

5 & Good & Good
& 0.266 & 0.0 & 8.2 & 95.8 & 2.9 & 3.64
& 0.1 & 8.2 & 95.6 & 2.9 & 3.70
& 0.99 [0.99, 1.00] & 0.02 \\
& Good & Limited
& 0.003 & 0.5 & 25.0 & 95.3 & 645.9 & 6.28
& $-$0.1 & 8.4 & 95.9 & 3.0 & 3.80
& 0.11 [0.05, 0.40] & 0.13 \\
& Limited & Good
& 0.266 & 0.0 & 11.6 & 95.0 & 2.9 & 3.87
& 0.1 & 11.5 & 94.7 & 2.9 & 3.81
& 0.99 [0.99, 1.00] & 0.02 \\
& Limited & Limited
& 0.003 & 0.2 & 19.4 & 95.0 & 18.6 & 7.19
& 0.2 & 12.4 & 93.9 & 3.1 & 4.15
& 0.41 [0.35, 0.50] & 0.13 \\
\bottomrule
\end{tabular}
\end{sidewaystable}
When follow-up overlap was good, CO with $c=0.1$ closely reproduced
OW. The variance ratio was 0.99 or 1.00 in all six such cells, and
the TV distance between the two target populations was at most 0.02.
Bias and coverage were also nearly identical. Thus, in these
configurations, incorporating follow-up information produced little
change in either the target population or precision when follow-up
overlap was adequate.

Under limited follow-up, the behavior was different. The sampling
distribution of OW became increasingly heavy-tailed as the horizon
increased, including under randomized treatment. In the randomized
setting, the variance ratio of CO relative to OW decreased from 0.89
(95\% Monte Carlo interval 0.85--0.93) at $\tau=1$ to 0.60
(0.53--0.66) at $\tau=3$ and 0.11 (0.05--0.40) at $\tau=5$.
When limited treatment and limited follow-up occurred together, the
corresponding ratios were 0.92, 0.53, and 0.41. The upper endpoint
of the Monte Carlo interval was below one in every limited-follow-up
configuration.

The tail summaries showed the same progression. Under randomized
treatment and limited follow-up, the OW tail index increased from
3.95 at $\tau=1$ to 4.83 at $\tau=3$ and 6.28 at $\tau=5$,
whereas the corresponding CO values remained between 3.80 and 4.01.
At $\tau=5$, the fifth percentile of $G_1(\tau\mid X)$ was 0.003,
and the empirical standard deviations were 0.250 for OW and 0.084
for CO. OW kurtosis was 646 in this cell, compared with 3.0 for CO.
The unusually wide Monte Carlo interval for the variance ratio
(0.05--0.40) provides a complementary indication of the same
instability: a small number of OW replicates had substantial influence
on its empirical variance. We therefore regard the direction of the
precision difference as clear in this configuration, while the exact
magnitude of the variance ratio is estimated less precisely.

The heavy tails did not translate into systematic undercoverage in
these experiments. Across all 12 cells, coverage ranged from 94.2\%
to 95.8\% for OW and from 93.9\% to 95.9\% for CO. Thus, at
$n=2000$, limited follow-up appeared primarily through instability
and loss of precision rather than failure of nominal Wald coverage.
This distinction is important: the simulations demonstrate the
finite-sample manifestation of the variance instability characterized
in Section~\ref{sec:finite}, not that OW intervals must fail whenever
follow-up probabilities are small.

The competing stabilization strategies involved different tradeoffs.
Under limited follow-up, the product tilt
$e(1-e)G_1G_0$ produced variance reductions similar to CO but moved
the target population farther from OW. Its TV distance was
0.28--0.36, compared with 0.08--0.13 for CO with $c=0.1$, or
2.8--3.6 times as large. When follow-up was good at $\tau=5$, the
product tilt instead had 8--9\% larger variance than OW.

Truncation retained the OW estimand but introduced a different
bias--variance tradeoff. Under limited follow-up at $\tau=5$,
truncating the estimated censoring survival function at its fifth
percentile reduced variance, with variance ratios 0.13 and 0.52, but
introduced bias of up to 0.078 relative to the OW target. At
$\tau=1$ under limited follow-up, the same rule increased variance,
with ratios 1.34 and 1.39. When follow-up was good, its coverage ranged from 90.3\% to 93.8\% at $\tau\in\{3,5\}$. Fixed truncation was less favorable at short horizons: under limited follow-up at $\tau=1$, its variance ratios ranged from 1.73 to 1.95. Full results for all truncation
rules are reported in Appendix~D.

The primary design also makes the cost of changing the target
explicit. Under randomized treatment and limited follow-up at
$\tau=5$, the true RMST contrast was 0.375 for CO with $c=0.1$ and
0.286 for OW. The difference arises because CO shifts weight away
from covariate profiles with weak follow-up, while the treatment
effect varies across those profiles. It is therefore a difference in
estimands rather than estimation bias. Reporting the target distance
and the $c$ path makes this change in the target population explicit.

\subsection{Additional experiments}\label{sec:simextra}

Appendix~D reports four auxiliary experiments designed to isolate
features of the primary experiment. These used 400 or 500 replicates
per configuration. Because the factorial study showed that
rare replicates can substantially affect the empirical tails of OW,
we use the auxiliary experiments for qualitative comparisons rather
than for precise characterization of tail behavior.

\emph{Constant target.}
We first removed target movement by allowing the outcome to depend
only on a covariate independent of the treatment and censoring
mechanisms, so that all tilts had the same estimand. Under limited
follow-up, CO reduced variance by factors of approximately 1.5 to 3,
corresponding to variance ratios between 0.33 and 0.65, without
changing the target. Thus, the stabilization observed in the primary
experiment does not depend on estimand movement.

\emph{Misjudged follow-up.}
We next deliberately misspecified the censoring model so that it
overstated follow-up beyond a covariate threshold. As the probability
of this region increased from 0 to 0.31, the bias of CO with $c=0.1$
changed from 0.003 to $-0.028$, while its empirical standard
deviation remained approximately 0.09. This experiment illustrates
the limitation represented by Condition~\ref{cond:V}: CO addresses
small follow-up probabilities when their location is estimated
adequately, but it cannot protect against persistent failure to
identify where follow-up is poor.

\emph{Alignment of treatment and follow-up information.}
We varied the alignment between covariate profiles carrying treatment
information and those with weak follow-up. CO with $c=0.1$ matched
or improved on the precision of OW in every configuration examined,
with variance ratios between 0.01 and 1.00. The smallest ratio
occurred when limited follow-up affected the population emphasized by
OW and OW became highly unstable. More aggressive tilts could instead
lose precision when retreating from weak follow-up removed
observations carrying substantial treatment information: with rare
treatment, variance increased by as much as 17\% for $c=1$ and 62\%
for the product tilt. These experiments illustrate why the regularity
results do not imply variance dominance of CO over OW.

\emph{Censoring-model sensitivity.}
Finally, we compared correctly specified, linear, and deliberately
flattened censoring models. Across the perturbations examined, the CO
estimate changed 5--10 times less than the OW estimate in absolute
terms and 1.7--3.4 times less relative to their standard deviations.
This reduced sensitivity is an empirical finding of the configurations
examined. Its direction accords with the cancellation underlying
Proposition~\ref{prop:remainder}, but that proposition does not imply
these finite-sample magnitudes.


\section{Application: adjuvant hormonal therapy in the Rotterdam cohort}
\label{sec:application}

\subsection{Data and analysis}\label{sec:rdata}

The Rotterdam tumour bank cohort \citep{royston2013external}, distributed
with the R package \texttt{survival}, contains 2982 women who underwent
primary surgery for breast cancer between 1978 and 1993. Overall survival
was followed for up to 19.3 years, with 1272 deaths. We compared adjuvant
hormonal therapy, received by 339 women (11\%), with no hormonal therapy.
Treatment use increased over calendar time, from 2\% in 1978--1982 to
4\% in 1983--1987 and 16\% in 1988--1993. Follow-up moved in the opposite
direction: among women alive at last follow-up, its correlation with year
of surgery was $-0.78$. The cohort therefore provides a setting in which
later patients are more likely to receive treatment but have less
long-term follow-up.

The estimand was the treated-minus-untreated difference in restricted
mean overall survival at $\tau\in\{5,7,8\}$ years. We excluded
$\tau=10$ before treatment-effect estimation because the fifth percentile
of $\hat G_a(10\mid X)$ was below 0.001 in both arms. The primary analysis
used CO weighting with $c=0.1$; OW, CO with $c\in\{0.3,1\}$, and the
product tilt $e(1-e)G_1G_0$ completed the prespecified comparison.

The propensity score was estimated by logistic regression with cubic
splines in year of surgery and age, menopausal status, tumour grade and
size, log numbers of positive nodes, progesterone and oestrogen receptor
levels, and chemotherapy. Arm-specific Cox models for the event and
censoring hazards used the same covariates. A year-stratified
Kaplan--Meier censoring estimator and a secondary chemotherapy comparison
were used for sensitivity analyses. The Kaplan--Meier estimator was
initially specified as the primary censoring model, but design diagnostics
obtained before examining treatment outcomes showed instability at
$\tau=8$ and its pooling of the treatment arms imposed $G_0=G_1$.
The arm-specific Cox model was therefore designated primary before the
outcome analysis; the bootstrap precision analysis was likewise specified
before it was run.

\subsection{Design diagnostics and results}\label{sec:rresults}

Treatment overlap was modest: 28\% of estimated propensity scores were in
$(0.1,0.9)$, with 469 women satisfying $0.2<\hat e(X)<0.8$. Among these
women, the proportion with $\hat G_1(\tau\mid X)<0.10$ increased from
0\% at 5 years to 9.6\% at 7 years and 34.3\% at 8 years; at 8 years,
27.5\% had $\hat G_1<0.05$. Correspondingly, the $c=0.1$ CO target
retained 99\%, 90\%, and 73\% of the OW target mass at 5, 7, and 8 years,
whereas the product tilt retained 80\%, 51\%, and 33\%.

\begin{table}[H]
\caption{Rotterdam cohort: restricted mean overall survival difference
(years), adjuvant hormonal therapy versus no hormonal therapy. SE and
95\% interval are influence-function based. Mass is target-population
mass relative to overlap weighting (OW); TV is total-variation distance
from the OW population. Bootstrap rows report the median ratio of squared
standard errors, CO ($c=0.1$) to OW, its 2.5th and 97.5th percentiles,
and the proportion exceeding one among 927 usable resamples.}
\label{tab:rotterdam}
\centering\footnotesize
\begin{tabular}{llrrlrr}
\toprule
$\tau$ & Target & $\hat\psi$ & SE & 95\% interval & Mass & TV\\
\midrule
5 & OW
  & 0.254 & 0.076 & (0.106, 0.402) & 1.00 & 0.00 \\
  & CO, $c=0.1$
  & 0.255 & 0.076 & (0.106, 0.404) & 0.99 & 0.01 \\
  & CO, $c=0.3$
  & 0.258 & 0.077 & (0.107, 0.409) & 0.96 & 0.02 \\
  & CO, $c=1$
  & 0.266 & 0.080 & (0.110, 0.422) & 0.89 & 0.05 \\
  & Product $e(1-e)G_1G_0$
  & 0.274 & 0.083 & (0.111, 0.438) & 0.80 & 0.09 \\
  & \multicolumn{6}{l}{\emph{Bootstrap:
  1.01 (1.01, 1.02); proportion above 1: 1.00}} \\
\addlinespace

7 & OW
  & 0.416 & 0.124 & (0.172, 0.659) & 1.00 & 0.00 \\
  & CO, $c=0.1$
  & 0.437 & 0.131 & (0.180, 0.694) & 0.90 & 0.05 \\
  & CO, $c=0.3$
  & 0.457 & 0.139 & (0.184, 0.730) & 0.80 & 0.11 \\
  & CO, $c=1$
  & 0.491 & 0.154 & (0.189, 0.792) & 0.64 & 0.19 \\
  & Product $e(1-e)G_1G_0$
  & 0.526 & 0.172 & (0.190, 0.863) & 0.51 & 0.29 \\
  & \multicolumn{6}{l}{\emph{Bootstrap:
  1.12 (1.05, 1.20); proportion above 1: 0.99}} \\
\addlinespace

8 & OW
  & 0.429 & 0.150 & (0.136, 0.722) & 1.00 & 0.00 \\
  & CO, $c=0.1$
  & 0.506 & 0.180 & (0.154, 0.858) & 0.73 & 0.18 \\
  & CO, $c=0.3$
  & 0.538 & 0.195 & (0.156, 0.919) & 0.61 & 0.25 \\
  & CO, $c=1$
  & 0.569 & 0.216 & (0.145, 0.993) & 0.46 & 0.33 \\
  & Product $e(1-e)G_1G_0$
  & 0.601 & 0.245 & (0.120, 1.081) & 0.33 & 0.43 \\
  & \multicolumn{6}{l}{\emph{Bootstrap:
  1.40 (1.06, 1.62); proportion above 1: 0.98}} \\
\bottomrule
\end{tabular}
\end{table}

At 5 years, OW and the primary CO analysis were nearly identical (Table~\ref{tab:rotterdam}):
$\hat\psi=0.254$ and 0.255 years, respectively, with SE 0.076 for both.
At 8 years, where follow-up overlap was substantially weaker, the
estimates separated along the protection path: 0.429 under OW, 0.506
under $c=0.1$, 0.538 under $c=0.3$, and 0.569 under $c=1$. The
corresponding CO target masses were 0.73, 0.61, and 0.46, with TV
distances 0.18, 0.25, and 0.33. The product tilt moved farther, retaining
33\% of the OW target mass with TV distance 0.43.

The primary CO analysis was less precise at the longer horizon. At
$\tau=8$, its SE was 0.180 compared with 0.150 for OW. Among 927 usable
bootstrap resamples, the median ratio of squared SEs was 1.40
(2.5th--97.5th percentiles 1.06--1.62), and the ratio exceeded one in
98\% of resamples. The variance decomposition localized this difference:
the event-martingale component was 0.0307 for CO and 0.0216 for OW,
against total variances 0.0323 and 0.0224, whereas the censoring component
associated with the moving target was only 0.00004. The treated-arm
effective sample size likewise decreased from 260 under OW to 206 under
CO. Thus, in this application the precision cost arose predominantly from
reweighting the event information rather than from estimating the
censoring-dependent target.

\subsection{Sensitivity analyses}\label{sec:rsens}

Under the year-stratified censoring estimator, the 8-year OW influence
function was substantially more concentrated: its kurtosis was 315 and
the ten largest contributions accounted for 45\% of its empirical
variance, compared with 32 and 24\% under CO. The corresponding estimates
were $0.259$ (SE $0.193$) for OW and 0.409 (SE 0.182) for CO. Switching
between the Cox and year-stratified censoring specifications changed the
hormonal-therapy estimate by 0.169 years under OW and 0.098 under CO; for
chemotherapy, the corresponding changes were $0.062$ and $0.038$.

The chemotherapy analysis also provided a setting with substantially
better follow-up overlap under the primary model: at 8 years, only 0.1\%
of women near treatment equipoise had $\hat G_1<0.05$. The OW and CO
estimates were correspondingly close, 0.441 (SE 0.129) and 0.455
(SE 0.135), with TV distance 0.06 between their target populations.


\section{Discussion}\label{sec:discussion}

Causal survival analyses require overlap not only in treatment assignment
but also in follow-up. We showed that these are distinct requirements:
even under randomized treatment, overlap-weighted RMST effects can have
an infinite semiparametric efficiency bound when follow-up probabilities
approach zero across covariate profiles. Censoring-adjusted overlap
weighting incorporates both mechanisms into a design-stage target. Its
efficient influence function is bounded without requiring
$G_a(\tau\mid X)$ to be uniformly bounded away from zero, while the same
structure controls inverse-censoring terms in the second-order remainder.
This yields regular inference under arbitrarily weak, but nonzero,
follow-up.

The empirical results reflect this distinction. CO with $c=0.1$ remained
close to OW when follow-up overlap was adequate and stabilized estimation
when follow-up was limited. The Rotterdam analysis also showed that this
protection can cost precision when poorly followed observations carry
substantial treatment information. CO is therefore not a
variance-dominance procedure, but a follow-up-aware target for settings
in which OW need not remain regular.

Three limitations define the scope of the method. First, CO changes the
target population when follow-up varies across covariate profiles, which
can matter under treatment-effect heterogeneity. We therefore recommend
reporting target mass, distance from the OW population, and the $c$ path.
This parallels the change in target induced by overlap weighting relative
to the average treatment effect \citep{li2018balancing,crump2009dealing}.
Truncating censoring weights instead nominally retains the original target
but can introduce bias, as in our simulations.

Second, protection requires the censoring model to identify where
follow-up is weak. Persistent misspecification in such regions can prevent
the intended down-weighting, so censoring-model diagnostics remain
important. The method also addresses limited rather than absent follow-up:
if $G_a(\tau\mid X)=0$ on a set of positive probability, a shorter horizon
or different target is required.

Third, the usual identification assumptions remain necessary: treatment
is unconfounded given baseline covariates, and censoring is conditionally
independent of event time given treatment and baseline covariates.
Conventional treatment positivity is also retained. CO removes the
analogous uniform lower-bound requirement on follow-up, but is not doubly
robust with respect to the treatment and censoring mechanisms defining
its target. Although the theory permits cross-fitted flexible nuisance
estimators, our numerical studies used finite-dimensional parametric and
Cox specifications.

The constant $c$ controls departure from the OW target. We used $c=0.1$
as the primary specification and $c\in\{0.3,1\}$ as a prespecified
protection path rather than selecting $c$ from the outcomes. Developing
outcome-free rules for choosing $c$, for example through an acceptable
loss of target mass, is a natural extension and connects to work on
estimand selection \citep{barnard2026framework}. Extensions to competing
risks, time-varying censoring mechanisms, and joint inference across
survival horizons are also of interest.

In practice, treatment and follow-up overlap should be examined jointly
before outcome analysis. At each horizon, the distribution of
$\hat G_a(\tau\mid X)$ among patients near treatment equipoise indicates
whether the population emphasized by treatment overlap is also supported
by follow-up. When support weakens, the CO protection path makes the
resulting target change explicit; when follow-up is structurally absent,
the horizon should instead be shortened. Treatment overlap alone is not
sufficient to establish support for a causal survival comparison.

\section*{Acknowledgements}
The authors are grateful to their supervisors for their guidance and
encouragement.
\section*{Disclosure statement}
The authors report no conflict of interest.
\section*{Funding}
No funding was received for this work.

\section*{Data Availability Statement}
The Rotterdam data used are publicly available in the
R package \texttt{survival} (data set \texttt{rotterdam}). All data and simulation code is available at \url{https://github.com/ntenkorang/causal-surv}.

\clearpage
\bibliographystyle{apalike}
\bibliography{references}



\newpage
\begin{appendix}

\begin{center}
    {\large \textbf{Supplementary Material for}}\\[0.1in]
    {\Large \textbf{``Limited Follow Up Overlap in Causal Survival Analysis: Censoring-Adjusted Weighting for Restricted Mean Survival Time''}} 
\end{center}

\maketitle
\sloppy

\noindent
Lemma~4 is stated here; other results refer to the numbering in the main text. Throughout, arm
$a$ and covariates $X$ are fixed whenever a statement concerns the
conditional law of $(\tilde T,\Delta)$ given $(A=a,X)$. We then write
$S=S_a(\cdot\mid X)$, $G=G_a(\cdot\mid X)$,
$\Lambda_T=\Lambda_{T,a}$, $\Lambda^C=\Lambda^C_a$,
$y(u)=S(u-)G(u-)$, $Y_\tau=\min(T,\tau)$, and
$\bar R(u)=\int_u^\tau S(t)\,dt$. Event and censoring laws are continuous
unless stated otherwise, so left limits may be dropped. We write
\[
N^T(u)=1(\tilde T\le u,\Delta=1),\qquad
N^C(u)=1(\tilde T\le u,\Delta=0),
\]
and
\[
dM^T=dN^T-1(\tilde T\ge u)\,d\Lambda_T,\qquad
dM^C=dN^C-1(\tilde T\ge u)\,d\Lambda^C.
\]

\section{Appendix A: results for Section 2}

\subsection{Proof of Lemma 1}

\paragraph{Step 1: an inverse-weight identity.}
Let $R_\tau=1(C\ge Y_\tau)$. For a continuous censoring law,
\begin{equation}\label{eq:A-ipcw}
\frac{R_\tau}{G(Y_\tau)}
=
1-\int_0^{Y_\tau}\frac{dM^C(u)}{G(u)}.
\end{equation}
On $[0,Y_\tau]$, the at-risk indicator for $T$ equals one, so only
censoring contributes. If $C\ge Y_\tau$, then $N^C$ does not jump before
$Y_\tau$ and
\[
\int_0^{Y_\tau}\frac{dM^C(u)}{G(u)}
=
-\int_0^{Y_\tau}\frac{d\Lambda^C(u)}{G(u)}
=
-\left\{\frac{1}{G(Y_\tau)}-1\right\}.
\]
Both sides of~\eqref{eq:A-ipcw} therefore equal $1/G(Y_\tau)$. If
$C<Y_\tau$, the integral equals
\[
\frac{1}{G(C)}
-
\left\{\frac{1}{G(C)}-1\right\}
=1,
\]
and both sides of~\eqref{eq:A-ipcw} are zero.

\paragraph{Step 2: decomposition.}
On $\{R_\tau=1\}$, $U=Y_\tau$; on $\{R_\tau=0\}$, the integral in transform~(2) stops at $U=C<Y_\tau$, beyond which $M^C$ is constant.
Hence
\[
\phi_a
=
\frac{R_\tau Y_\tau}{G(Y_\tau)}
+
\int_0^{Y_\tau}\frac{m(u)}{G(u)}\,dM^C(u).
\]
Substituting~\eqref{eq:A-ipcw} gives
\[
\phi_a=Y_\tau-W_a,
\qquad
W_a=
\int_0^{Y_\tau}
\frac{Y_\tau-m(u)}{G(u)}\,dM^C(u).
\]

\paragraph{Step 3: orthogonality.}
Condition on $T$. The upper limit $Y_\tau$ is then fixed, and the
integrand is deterministic and bounded on $[0,Y_\tau]$ under
Condition~1. Because $C\perp\!\!\!\perp T\mid(A,X)$,
\[
N^C(u)-\int_0^u1(C\ge v)\,d\Lambda^C(v),
\qquad u\le Y_\tau,
\]
is a martingale in the filtration generated by $C$. Thus
$E(W_a\mid T)=0$, so $E\{g(T)W_a\}=0$ for every square-integrable
$g$. Consequently,
\[
v_a
=
\operatorname{var}(Y_\tau)+E(W_a^2)
=
\sigma_a^2+E(W_a^2).
\]

\paragraph{Step 4: variance of $W_a$.}
By predictable variation of the censoring martingale conditional on
$T$, and using $\Pr(C\ge u)=G(u)$,
\[
E(W_a^2\mid T)
=
\int_0^{Y_\tau}
\frac{\{Y_\tau-m(u)\}^2}{G(u)}\,d\Lambda^C(u).
\]
For $u\le\tau$, $\{u\le Y_\tau\}=\{T\ge u\}$ and
\[
E[\{Y_\tau-m(u)\}^2\mid T\ge u]
=
\omega_a(u,X).
\]
Taking expectations and applying Fubini's theorem gives
\[
E(W_a^2)
=
\int_0^\tau
\omega_a(u,X)S(u-)
\frac{d\Lambda^C(u)}{G(u)}
=
C_a(X).
\]

\paragraph{Step 5: bounds.}
Conditional on $T\ge u$, $Y_\tau\in[u,\tau]$, and therefore
\[
\omega_a(u,X)
\le
\frac{(\tau-u)^2}{4}
\le
\frac{\tau^2}{4}.
\]
Since $S\le1$ and
\[
\int_0^\tau\frac{d\Lambda^C(u)}{G(u)}
=
\frac{1}{G(\tau)}-1,
\]
the upper bound follows. For the lower bound, restrict the integral to
$[0,s]$, use $S(u-)\ge S(s)$ and
\[
\omega_a(u,X)
\ge
\inf_{v\le s}\omega_a(v,X),
\]
and note that
\[
\int_0^s\frac{d\Lambda^C(u)}{G(u)}
=
\frac{1}{G(s)}-1.
\]
\qed

\paragraph{Remark on atoms.}
If the censoring law has atoms, Step~1 uses $G(u-)$ in the denominators
and
\[
d\{1/G(u)\}=\frac{d\Lambda^C(u)}{G(u)}.
\]
Since $G(u-)\ge G(u)$, the upper bound in Step~5 remains valid. The lower
bound above is stated for the continuous case.


\subsection{Proof of Theorem 1}

Consider bounded tilts $0\le h\le\bar h$ with $E(h)>0$, which include
all tilts considered in the paper. By~(4),
\[
E(h)^2V(h)
=
\sum_a
E\left\{\frac{h^2v_a}{e_a}\right\}
+
E\{h^2(\Delta-\psi_h)^2\},
\]
and the last term is at most $4\tau^2\bar h^2$.

\paragraph{Part (i).}
By Lemma~1,
\[
v_a
\le
\frac{\tau^2}{4}
+
\frac{\tau^2}{4}
\left\{\frac{1}{G_a(\tau\mid X)}-1\right\}
=
\frac{\tau^2}{4G_a(\tau\mid X)}.
\]
Therefore
\[
E\left(\frac{h^2v_a}{e_a}\right)
\le
\frac{\tau^2}{4}
E\left\{
\frac{h^2}
{e_aG_a(\tau\mid X)}
\right\}
<\infty.
\]

\paragraph{Part (ii).}
Under Condition~2,
\[
S_a(s\mid X)\ge S_a(\tau\mid X)\ge s_0.
\]
Condition~3 at $u=0$ gives
$\sigma_a^2=\omega_a(0,X)\ge\omega_{\min}$, and for $u\le s$,
$\omega_a(u,X)\ge\omega_{\min}$.

If $E(h^2/e_a)=\infty$, then
\[
E\left(\frac{h^2v_a}{e_a}\right)
\ge
\omega_{\min}E\left(\frac{h^2}{e_a}\right)
=
\infty.
\]
Otherwise, the lower bound in Lemma~1 gives
\[
E\left(\frac{h^2v_a}{e_a}\right)
\ge
s_0\omega_{\min}
\left[
E\left\{
\frac{h^2}{e_aG_a(s\mid X)}
\right\}
-
E\left(\frac{h^2}{e_a}\right)
\right]
=
\infty.
\]
Thus $V(h)=\infty$. \qed


\subsection{Proof of Corollary 1 and the beta-family extension}

\paragraph{Conditions.}
With event hazard $\lambda>0$,
\[
S_a(\tau\mid X)=e^{-\lambda\tau},
\]
so Condition~2 holds. By the memoryless property,
\[
\omega_a(u,X)
=
\operatorname{var}\{\min(T',\tau-u)\},
\qquad
T'\sim\operatorname{Exp}(\lambda),
\]
which is continuous and strictly positive on the compact interval
$[0,s]$. Hence Condition~3 holds. Finally,
\[
G_a(\tau\mid X)
=
\exp\{-\lambda_C\tau e^{\gamma X}\}>0,
\]
and $0<p<1$, so Condition~1 holds.

\paragraph{Divergence.}
With $e\equiv p$, both $h_{\OW}=p(1-p)$ and $h\equiv1$ are positive
constants. By Theorem~1(ii), it is therefore enough to show that
\[
E\left\{\frac{1}{G_a(s\mid X)}\right\}
=
E\left[
\exp\{\lambda_Cs e^{\gamma X}\}
\right]
=
\infty.
\]
Take $\gamma>0$; the case $\gamma<0$ is symmetric. For $x\ge0$, the
integrand in the expectation is proportional to
\[
\exp\left\{
\lambda_Cs e^{\gamma x}-\frac{x^2}{2}
\right\},
\]
which tends to infinity because $e^{\gamma x}$ dominates $x^2$.
Therefore the integral diverges.

\paragraph{Beta family.}
Let
\[
\operatorname{logit}e(X)=\alpha_0+\alpha_1X,
\qquad
h=\{e(1-e)\}^{\nu},
\qquad
\nu\ge0.
\]
Writing $\eta=\alpha_0+\alpha_1X$,
\[
e(1-e)
=
\frac{e^\eta}{(1+e^\eta)^2}
\ge
\frac{e^{-|\eta|}}{4}.
\]
Since $e_a\le1$,
\[
\frac{h^2}{e_aG_a(s\mid X)}
\ge
4^{-2\nu}
\exp\{
\lambda_Cs e^{\gamma X}
-
2\nu|\alpha_0+\alpha_1X|
\}.
\]
After multiplication by the normal density, the exponent is
\[
\lambda_Cs e^{\gamma x}
-
2\nu|\alpha_0+\alpha_1x|
-
\frac{x^2}{2},
\]
which tends to infinity as $\gamma x\to\infty$. Hence the expectation
is infinite, and Theorem~1(ii) gives $V(h)=\infty$ for every
$\nu\ge0$. \qed


\subsection{The bounded-support sequence}

Let $X$ be standard normal truncated to $|X|<b$, with event hazard
$0.25$, censoring hazard $0.05e^X$ in both arms, and $\tau=4$. The
treatment effect is zero, so only the first term of~(4) contributes.
Table~\ref{tab:A-bounded} reports $V(h)$ for overlap weighting and CO
weighting with $c=0.1$. Values were computed by adaptive quadrature with
relative tolerance $10^{-10}$ in both $u$ and $x$.

\begin{table}[h]
\caption{Asymptotic variance under truncated covariate support for
overlap weighting (OW) and CO weighting with $c=0.1$. Mass is
$E(h_c)/E(h_{\OW})$.}
\label{tab:A-bounded}
\centering
\begin{tabular}{r rrr rrr}
\toprule
&
\multicolumn{3}{c}{Randomized, $e\equiv1/2$}
&
\multicolumn{3}{c}{Logistic, $\operatorname{logit}e=0.5X$}
\\
\cmidrule(lr){2-4}\cmidrule(lr){5-7}
$b$
& $V(h_{\OW})$ & $V(h_c)$ & Mass
& $V(h_{\OW})$ & $V(h_c)$ & Mass\\
\midrule
2 & 8.91 & 8.87 & 0.964 & 9.31 & 9.28 & 0.965 \\
3 & 9.04 & 8.94 & 0.957 & 9.51 & 9.42 & 0.959 \\
4 & 9.16 & 8.95 & 0.956 & 9.59 & 9.44 & 0.959 \\
5 & $1.04\times10^{3}$ & 8.95 & 0.955
  & $3.41\times10^{2}$ & 9.44 & 0.959 \\
6 & $8.49\times10^{20}$ & 8.95 & 0.955
  & $1.73\times10^{20}$ & 9.44 & 0.959 \\
\bottomrule
\end{tabular}
\end{table}

Every law in the sequence satisfies Conditions~1--3 with the same
event-time constants. The target mass is bounded below uniformly in $b$:
$E(h_c)\ge0.2389$ in the randomized scenario and
$E(h_c)\ge0.2263$ in the logistic scenario, with the minima attained for
$b\ge4$. The laws within each scenario therefore satisfy the corresponding
fixed-constant requirements of Theorem~5 uniformly in $b$.

\section{Appendix B: results for Sections 3--5}

\subsection{Proof of Proposition 1}

Write
\[
\kappa_{a,c}
=
1-c+\frac{c}{G_a(\tau\mid X)},
\qquad
c\in(0,1],
\qquad
G_a(\tau\mid X)\in(0,1].
\]

\paragraph{Part (i).}
If $G_a(\tau\mid X)=1$, then $\kappa_{a,c}=1$ and
\[
h_c
=
\left\{\frac1e+\frac1{1-e}\right\}^{-1}
=
e(1-e).
\]
For fixed $G_a$, $\kappa_{a,c}\to1$ as $c\downarrow0$, and $h_c$ is
continuous in $(\kappa_{1,c},\kappa_{0,c})$.

\paragraph{Part (ii).}
Since
\[
\frac{\partial\kappa_{a,c}}{\partial c}
=
\frac1{G_a}-1
\ge0,
\]
and $h_c$ is decreasing in each $\kappa_{a,c}$, increasing $c$
increases the degree of follow-up protection.

\paragraph{Part (iii).}
Both terms in $1/h_c$ are positive, so
\[
h_c\le\frac{e_a}{\kappa_{a,c}}
\]
for each $a$. Since $\kappa_{a,c}\ge1$,
\[
h_c\le\min(e,1-e)\le1.
\]
Also,
\[
\kappa_{a,c}\ge\frac{c}{G_a(\tau\mid X)},
\]
which gives
\[
h_c
\le
\frac{e_aG_a(\tau\mid X)}{c}.
\]
Positivity follows from Condition~1.

\paragraph{Part (iv).}
By Lemma~1,
\[
v_a
\le
\frac{\tau^2}{4G_a(\tau\mid X)}.
\]
Combining this with
$\kappa_{a,c}\ge c/G_a(\tau\mid X)$ gives
\[
\frac{v_a}{\kappa_{a,c}}
\le
\frac{\tau^2}{4c}.
\]
\qed


\subsection{Proof of Lemma 2}

In the conditional model for $(\tilde T,\Delta)$ given $(A=a,X=x)$,
the law is determined by $(\Lambda_T,\Lambda^C)$, and every mean-zero
square-integrable function has the representation
\[
\int k_T\,dM^T+\int k_C\,dM^C
\]
\citep[Chapters 9--10]{tsiatis2006semiparametric}. The conditional model
is therefore locally nonparametric, so a gradient, once identified, is
canonical.

Consider the censoring-direction submodel
\[
d\Lambda_t^C=(1+tk)\,d\Lambda^C,
\]
with bounded $k$. Its score is $\int k\,dM^C$, and
\[
\left.
\frac{\partial}{\partial t}\Lambda_t^C(\tau)
\right|_{t=0}
=
\int_0^\tau k\,d\Lambda^C.
\]
By predictable covariation and
$E\{1(\tilde T\ge u)\}=y(u)$,
\[
E\left\{
\int_0^\tau\frac{dM^C}{y}
\int k\,dM^C
\right\}
=
\int_0^\tau
\frac{k(u)}{y(u)}y(u)\,d\Lambda^C(u)
=
\int_0^\tau k\,d\Lambda^C.
\]
Along event-direction submodels, $\Lambda^C$ does not change, and
\[
E\left\{
\int_0^\tau\frac{dM^C}{y}
\int k\,dM^T
\right\}
=0,
\]
because $\langle M^C,M^T\rangle=0$ for continuous event and censoring
laws without common jumps. Hence
\[
\int_0^\tau\frac{dM^C(u)}{y(u)}
\]
is the gradient of $\Lambda^C(\tau\mid x)$. Since
$G=\exp(-\Lambda^C)$, the chain rule gives
\[
\gamma_a
=
-G(\tau\mid x)
\int_0^\tau\frac{dM^C(u)}{y(u)}.
\]
\qed


\subsection{An identity for the AIPCW transform}

\begin{lemma}\label{lem:B-identity}
Let $\tilde S$ and $\tilde G$ be absolutely continuous survival
functions on $[0,\tau]$, with cumulative hazards
$\tilde\Lambda_T$ and $\tilde\Lambda^C$. Define
\[
\tilde R(u)=\int_u^\tau\tilde S(t)\,dt,
\qquad
\tilde\mu=\tilde R(0),
\qquad
\tilde m(u)=u+\frac{\tilde R(u)}{\tilde S(u)},
\]
and let $\tilde\phi$ denote transform~(2) constructed from
$(\tilde S,\tilde G)$. Then, for every observation,
\[
\tilde\phi
=
\tilde\mu
-
\int_0^U
\frac{\tilde R(u)}
{\tilde S(u)\tilde G(u)}
\left\{
dN^T(u)
-
1(\tilde T\ge u)\,d\tilde\Lambda_T(u)
\right\}.
\]
\end{lemma}

\begin{proof}
Differentiating $\tilde m/\tilde G$ and using
\[
d\tilde R=-\tilde S\,du,\qquad
d\tilde S=-\tilde S\,d\tilde\Lambda_T,\qquad
d(1/\tilde G)=d\tilde\Lambda^C/\tilde G,
\]
gives
\[
d\left(\frac{\tilde m}{\tilde G}\right)
=
\frac{\tilde m\,d\tilde\Lambda^C}{\tilde G}
+
\frac{\tilde R\,d\tilde\Lambda_T}
{\tilde S\tilde G}.
\]
Since $\tilde m(0)=\tilde\mu$,
\[
\int_0^U
\frac{\tilde m\,d\tilde\Lambda^C}{\tilde G}
=
\frac{\tilde m(U)}{\tilde G(U)}
-
\tilde\mu
-
\int_0^U
\frac{\tilde R\,d\tilde\Lambda_T}
{\tilde S\tilde G}.
\]

If the observation is censored before $\tau$, then $R_\tau=0$ and
$U=C$. If an event occurs before $\tau$, then $U=T$ and
\[
\tilde m(U)-U=\frac{\tilde R(U)}{\tilde S(U)}.
\]
If the subject is followed through $\tau$, then $U=\tau$ and
$\tilde m(\tau)=\tau$. Substitution in each of these three cases yields
the stated representation; the jump in $N^T$ appears only when an event
occurs before $\tau$.
\end{proof}

With the true nuisances, Lemma~\ref{lem:B-identity} gives the
event-martingale representation of $\phi_a-\mu_a$ used in Section~4.2.


\subsection{Proof of Theorem 3}

Let $P_t$ be a regular one-dimensional submodel with bounded score
\[
s=s_X+s_A+s_Y,
\]
where $E(s_A\mid X)=0$ and $E(s_Y\mid A,X)=0$. Write
\[
s_Y=s_T+s_C,
\qquad
s_T=\int k_T\,dM^T,
\qquad
s_C=\int k_C\,dM^C.
\]
Let $h_t$, $\Delta_t$, and $\psi_t$ denote the corresponding quantities
under $P_t$, and use a dot for differentiation at $t=0$.
Differentiating
\[
\psi_t=\frac{E_t(h_t\Delta_t)}{E_t(h_t)}
\]
gives
\begin{equation}\label{eq:B-ratio}
E(h)\dot\psi
=
E\{s_Xh(\Delta-\psi)\}
+
E\{\dot h(\Delta-\psi)\}
+
E(h\dot\Delta),
\end{equation}
because $E\{h(\Delta-\psi)\}=0$.

\paragraph{Covariate term.}
Since $h(\Delta-\psi)$ is a mean-zero function of $X$,
\[
E\{s_Xh(\Delta-\psi)\}
=
E\{s\,h(\Delta-\psi)\}.
\]

\paragraph{Outcome term.}
The contrast $\Delta$ depends on $P$ only through the event hazards.
In the conditional model, $\mu_a(x)$ has gradient $\epsilon_a=\phi_a-\mu_a$
which is orthogonal to censoring scores by
Lemma~\ref{lem:B-identity}. Thus
\[
\dot\mu_a(x)
=
E(\epsilon_a s_Y\mid A=a,X=x).
\]
For any bounded $g(X)$,
\[
E\{g\dot\mu_a\}
=
E\left\{
g\frac{1(A=a)}{e_a}\epsilon_a s
\right\},
\]
because $E\{1(A=a)\epsilon_a\mid X\}=0$ eliminates the $s_X$ and $s_A$
terms.Taking $g=h$ gives the outcome term in~(6).

\paragraph{Tilt term.}
The tilt depends on $P$ through $e$ and
$G_a(\tau\mid\cdot)$:
\[
\dot h
=
H_e\dot e
+
\sum_aH_{g_a}\dot G_a(\tau\mid X).
\]
For the propensity score,
\[
\dot e(x)
=
E\{(A-e)s_A\mid X=x\},
\]
so
\[
E\{(\Delta-\psi)H_e\dot e\}
=
E\{(\Delta-\psi)H_e(A-e)s\}.
\]
For censoring, Lemma~2 gives
\[
\dot G_a(\tau\mid x)
=
E(\gamma_a s_C\mid A=a,X=x),
\]
and hence
\[
E\{(\Delta-\psi)H_{g_a}\dot G_a\}
=
E\left\{
(\Delta-\psi)H_{g_a}
\frac{1(A=a)}{e_a}\gamma_a s
\right\}.
\]
Substitution of
\[
H_{g_a}
=
\frac{ch^2}{e_aG_a^2}
\]
and the expression for $\gamma_a$ gives the final term of~(6).

The event and censoring contributions occupy orthogonal martingale
spaces, so each perturbation direction is represented once. Collecting
terms gives
\[
\dot\psi=E(D_{\CO}s),
\]
where $D_{\CO}$ is given by~(6). Proposition~2 shows that $D_{\CO}$ is
bounded and therefore belongs to $L_2(P)$. Since the observed-data model
is locally nonparametric, this gradient is canonical
\citep{gill1997coarsening}. \qed


\subsection{Proof of Corollary 2}

Write
\[
D_{\CO}
=
\frac{D_X+D_A+D_T+D_C}{E(h)}
\]
for the four components of~(6). The components are pairwise orthogonal:
$D_X$ is a function of $X$ and $E(D_A\mid X)=0$; both $D_T$ and $D_C$
have conditional mean zero given $(A,X)$; and, by
Lemma~\ref{lem:B-identity},
\[
E\left\{
\int k_T\,dM^T
\int k_C\,dM^C
\Bigm|A,X
\right\}
=
E\int k_Tk_C\,d\langle M^T,M^C\rangle
=
0.
\]

The corresponding conditional second moments are
\[
E(D_A^2\mid X)
=
(\Delta-\psi_c)^2H_e^2e(1-e),
\]
\[
E(D_T^2\mid X)
=
h_c^2\sum_a\frac{v_a}{e_a},
\]
and, using
\[
E\left\{
\left(
\int_0^\tau\frac{dM_a^C}{y_a}
\right)^2
\Bigm|A=a,X
\right\}
=
\int_0^\tau\frac{d\Lambda_a^C}{y_a},
\]
\[
E(D_C^2\mid X)
=
c^2(\Delta-\psi_c)^2h_c^4
\sum_a
\frac{1}
{e_a^3G_a(\tau\mid X)^2}
\int_0^\tau\frac{d\Lambda_a^C}{y_a}.
\]
\qed


\subsection{Proof of Proposition 2}

We bound the four terms of~(6), multiplied by $E(h_c)$, pointwise.
Throughout,
\[
|\Delta-\psi_c|\le2\tau.
\]

\paragraph{Covariate term.}
Since $h_c\le1$,
\[
|h_c(\Delta-\psi_c)|\le2\tau.
\]

\paragraph{Treatment term.}
We have
\[
H_e
=
\frac{h_c^2\kappa_{1,c}}{e^2}
-
\frac{h_c^2\kappa_{0,c}}{(1-e)^2}.
\]
Each component lies in $[0,1]$, because
\[
\frac{h_c\kappa_{a,c}}{e_a}\le1,
\qquad
\frac{h_c}{e_a}\le1.
\]
Hence $|H_e|\le1$, and the treatment term is bounded by $2\tau$.

\paragraph{Outcome term.}
Only one treatment arm is active for a given observation. The first two
terms of~(6) cannot both be nonzero, and each is bounded by
$\tau/G_a(U)$. The compensator is bounded by
\[
\tau\left\{\frac1{G_a(U)}-1\right\}.
\]
Therefore
\[
|\phi_a|\le\frac{\tau}{G_a(U)}.
\]
By Proposition~1(iii),
\[
\left|
\frac{h_c\phi_a}{e_a}
\right|
\le
\frac{G_a(\tau)}{c}
\frac{\tau}{G_a(U)}
\le
\frac{\tau}{c},
\]
since $U\le\tau$. Also,
\[
\frac{h_c\mu_a}{e_a}\le\tau.
\]
Thus the outcome term is bounded by
\[
\frac{\tau}{c}+\tau.
\]

\paragraph{Censoring term.}
Since $y_a\ge s_0G_a$, the jump of
$\int dM_a^C/y_a$ is bounded by
$1/\{s_0G_a(U)\}$, and its compensator is bounded by the same quantity.
Hence
\[
|\gamma_a|
\le
\frac{G_a(\tau)}
{s_0G_a(U)}
\le
\frac1{s_0}.
\]
Moreover,
\[
\frac{H_{g_a}}{e_a}
=
\frac{ch_c^2}
{e_a^2G_a(\tau)^2}
\le
\frac1c
\]
by Proposition~1(iii). The censoring term is therefore bounded by
\[
\frac{2\tau}{cs_0}.
\]

Summing the four bounds gives
\[
\|D_{\CO}\|_\infty
\le
\frac{\tau}{E(h_c)}
\left(
5+\frac1c+\frac{2}{cs_0}
\right).
\]

The same argument applies to $D_{\CO,\hat P}$ when
$\hat S_a(\tau\mid x)\ge s_0/2$. It uses
\[
\hat h
\le
\frac{\hat e_a\hat G_a(\tau)}{c},
\]
which holds by construction, with $s_0$ replaced by $s_0/2$. \qed


\subsection{Proofs of Lemma 3 and the expected censoring gradient}

\paragraph{Lemma 3.}
Apply Lemma~\ref{lem:B-identity} with
$(\tilde S,\tilde G)=(\hat S_a,\hat G_a)$. Under $P$,
\[
E\{dN^T(u)\mid A=a,X\}
=
S(u)G(u)\,d\Lambda_T(u)
\]
and
\[
\Pr(\tilde T\ge u\mid A=a,X)
=
S(u)G(u).
\]
Therefore
\[
E_P(\hat\phi_a\mid A=a,X)
=
\hat\mu_a
-
\int_0^\tau
\frac{\hat R(u)}
{\hat S(u)\hat G(u)}
S(u)G(u)
\,d(\Lambda_T-\hat\Lambda_T)(u).
\]

Duhamel's identity gives
\[
\hat S(t)-S(t)
=
\hat S(t)
\int_0^t
\frac{S(u)}{\hat S(u)}
\,d(\Lambda_T-\hat\Lambda_T)(u).
\]
Integrating over $t\in[0,\tau]$ and exchanging the order of integration,
\[
\hat\mu_a-\mu_a
=
\int_0^\tau
\hat R(u)
\frac{S(u)}{\hat S(u)}
\,d(\Lambda_T-\hat\Lambda_T)(u).
\]
Subtracting yields
\[
r_a
=
\int_0^\tau
\hat R(u)
\frac{S(u)}{\hat S(u)}
\left\{
1-\frac{G(u)}{\hat G(u)}
\right\}
\,d(\Lambda_T-\hat\Lambda_T)(u).
\]
\qed

\paragraph{Expected censoring gradient.}
Let
\[
\hat\gamma_a
=
-\hat G(\tau)
\int_0^\tau
\frac{d\hat M^C(u)}{\hat y(u)},
\]
where
\[
d\hat M^C
=
dN^C
-
1(\tilde T\ge u)\,d\hat\Lambda^C.
\]
The corresponding conditional expectation is
\begin{equation}\label{eq:B-L}
L_a(X)
=
E_P(\hat\gamma_a\mid A=a,X)
=
-\hat G(\tau)
\int_0^\tau
\frac{S(u)G(u)}
{\hat S(u)\hat G(u)}
\,d(\Lambda^C-\hat\Lambda^C)(u).
\end{equation}


\subsection{Proof of Proposition 3}

\paragraph{Exact decomposition.}
Let $\hat P$ share the covariate law of $P$, write $\hat\psi$ for
$\psi_c(\hat P)$, and define
\[
b_a(X)
=
E_P(\hat\phi_a-\hat\mu_a\mid A=a,X)
=
\mu_a-\hat\mu_a+r_a.
\]
Since
\[
E\{\hat h(\hat\Delta-\hat\psi)\}=0,
\]
we have
\[
\begin{split}
E(\hat h)P D_{\CO,\hat P}
={}&
E\left[
\hat h
\left\{
\frac{e}{\hat e}b_1
-
\frac{1-e}{1-\hat e}b_0
\right\}
\right] \\
&+
E\{(\hat\Delta-\hat\psi)\hat H_e(e-\hat e)\}\\
&+
\sum_a
E\left\{
(\hat\Delta-\hat\psi)
\hat H_{g_a}
\frac{e_a}{\hat e_a}
L_a
\right\}.
\end{split}
\]
Also,
\[
E(\hat h)(\hat\psi-\psi)
=
E\{\hat h(\hat\Delta-\Delta)\}
+
E\{(\hat h-h)(\Delta-\psi)\}.
\]
Combining the two expressions and writing
\[
\hat\Delta-\hat\psi
=
(\Delta-\psi)
+
\{(\hat\Delta-\Delta)-(\hat\psi-\psi)\}
\]
gives
\[
E(\hat h)R_2
=
\mathcal O
+
E\{(\Delta-\psi)\rho(X)\}
+
\mathcal X,
\]
where
\begin{align*}
\mathcal O
&=
E\left[
\hat h
\left\{
(\hat e-e)
\left(
\frac{\hat\mu_1-\mu_1}{\hat e}
+
\frac{\hat\mu_0-\mu_0}{1-\hat e}
\right)
+
\frac{e}{\hat e}r_1
-
\frac{1-e}{1-\hat e}r_0
\right\}
\right],\\
\rho
&=
\hat h-h
+
\hat H_e(e-\hat e)
+
\sum_a
\hat H_{g_a}
\frac{e_a}{\hat e_a}L_a,\\
\mathcal X
&=
E\left[
\{(\hat\Delta-\Delta)-(\hat\psi-\psi)\}
\left\{
\hat H_e(e-\hat e)
+
\sum_a
\hat H_{g_a}
\frac{e_a}{\hat e_a}L_a
\right\}
\right].
\end{align*}

\paragraph{Two weight bounds.}
By construction,
\[
\hat h
\le
\frac{\hat e_a\hat G_a(\tau)}{c},
\qquad
\hat S_a\ge\frac{s_0}{2}.
\]
Thus, for $u\le\tau$,
\begin{equation}\label{eq:B-weights}
\frac{\hat h}
{\hat e_a\hat G_a(u)}
\le
\frac1c,
\qquad
\hat H_{g_a}\hat G_a(\tau)
\frac{G_a(u)}{\hat G_a(u)}
=
\frac{c\hat h^2}
{\hat e_a\hat G_a(\tau)}
\frac{G_a(u)}{\hat G_a(u)}
\le
\frac{\hat e_aG_a(u)}{c}.
\end{equation}

\paragraph{Outcome term.}
Since $\hat R\le\tau$ and $S/\hat S\le2/s_0$,
Lemma~3 and~\eqref{eq:B-weights} give
\[
\left|
\hat h
\frac{e_a}{\hat e_a}
r_a
\right|
\le
\frac{2\tau}{cs_0}
\int_0^\tau
|\hat G_a-G_a|
\,|d(\Lambda_{T,a}-\hat\Lambda_{T,a})|.
\]
Also $\hat h/\hat e_a\le1$. Hence
\[
|\mathcal O|\le C\rho_1.
\]

\paragraph{Tilt term: global bound.}
We have
\[
|\hat h-h|\le2,
\qquad
|\hat H_e(e-\hat e)|\le1.
\]
From~\eqref{eq:B-L}, using
\[
\int_0^\tau G\,d\Lambda^C\le1
\]
and
\[
\int_0^\tau
\frac{d\hat\Lambda^C}{\hat G}
=
\frac1{\hat G(\tau)}-1,
\]
we obtain
\[
|L_a|
\le
\frac{2}{s_0}\hat G_a(\tau)
\left\{
\int_0^\tau
\frac{G_a\,d\Lambda_a^C}{\hat G_a}
+
\int_0^\tau
\frac{d\hat\Lambda_a^C}{\hat G_a}
\right\}
\le
\frac4{s_0}.
\]
Therefore
\[
\left|
\hat H_{g_a}
\frac{e_a}{\hat e_a}
L_a
\right|
\le
\frac4{cs_0},
\]
and
\[
|\rho|
\le
3+\frac8{cs_0}.
\]

\paragraph{Tilt term: local bound.}
On
\[
\mathcal G_M
=
\{|\delta_1|\le M,\ |\delta_0|\le M\},
\]
write
\[
\rho
=
-\operatorname{Rem}_H
+
\sum_a
\hat H_{g_a}
\{(\hat G_a-G_a)+L_a\}
+
\sum_a
\hat H_{g_a}
\frac{e_a-\hat e_a}{\hat e_a}
L_a,
\]
where
\[
\operatorname{Rem}_H
=
h-\hat h
-
\hat H_e(e-\hat e)
-
\sum_a
\hat H_{g_a}(G_a-\hat G_a)
\]
is the second-order Taylor remainder.

Work in coordinates
\[
(e,\ell_1,\ell_0),
\qquad
\ell_a=\log g_a.
\]
Along the segment between
$(\hat e,\hat\ell)$ and $(e,\ell)$,
\[
g_a\le e^MG_a,
\qquad
e\in[\epsilon_e/2,1-\epsilon_e/2].
\]
The derivative bounds of Appendix~C.2 give
\[
|H_{\ell_a\ell_a}|
\le
\frac{3e^MG_a}{c},
\qquad
|H_{e\ell_a}|
\le
\frac{6e^MG_a}{c\epsilon_e},
\qquad
|H_{ee}|
\le
\frac{C}{\epsilon_e}.
\]
Converting the linear term from $g$ to $\ell$ contributes at most
\[
\frac12
|\hat H_{\ell_a}|e^M\delta_a^2,
\qquad
|\hat H_{\ell_a}|
\le
\frac{e^MG_a}{c}.
\]
Applying $2|xy|\le x^2+y^2$ to the cross terms gives
\[
|\operatorname{Rem}_H|
\le
C(M)
\left\{
(\hat e-e)^2
+
\sum_aG_a(\tau)\delta_a^2
\right\}.
\]

Duhamel's identity for $G_a$ and~\eqref{eq:B-L} give
\[
(\hat G_a-G_a)+L_a
=
\hat G_a(\tau)
\int_0^\tau
\frac{G_a}{\hat G_a}
\frac{\hat S_a-S_a}{\hat S_a}
\,d(\Lambda_a^C-\hat\Lambda_a^C).
\]
By~\eqref{eq:B-weights}, this contribution is bounded by
\[
\frac{2}{cs_0}
\int_0^\tau
G_a
|\hat S_a-S_a|
\,|d(\hat\Lambda_a^C-\Lambda_a^C)|.
\]
Similarly,
\[
|\hat H_{g_a}L_a|
\le
\frac{2\hat e_a}{cs_0}
\int_0^\tau
G_a
\,|d(\Lambda_a^C-\hat\Lambda_a^C)|,
\]
so the propensity-by-censoring contribution is bounded by
\[
\frac{2}{cs_0}
|\hat e-e|
\int_0^\tau
G_a
\,|d(\hat\Lambda_a^C-\Lambda_a^C)|.
\]

Off $\mathcal G_M$, use the global bound. Since
$|\Delta-\psi|\le2\tau$,
\[
|E\{(\Delta-\psi)\rho\}|
\le
C(\rho_2+\rho_3+\rho_4),
\]
where the propensity-by-censoring product is included in the second
part of $\rho_3$.

\paragraph{Cross term.}
Using $|\hat H_e|\le1$ and the preceding bound on
$|\hat H_{g_a}L_a|$,
\[
|\mathcal X|\le C\rho_5.
\]

\paragraph{Normalization.}
Under Condition~5,
\[
\hat h\to h
\]
in probability at a new covariate value. Since both are bounded by one,
bounded convergence gives
\[
E|\hat h-h|\to0.
\]
Hence
\[
E(\hat h)\ge\frac{m_0}{2}
\]
with probability tending to one. Combining the preceding bounds proves
the proposition, with $C$ depending on
$(\tau,c,s_0,\epsilon_e,m_0,M)$. \qed


\subsection{Proof of Theorem 4}

Write
\[
\hat D=D_{\CO,\hat P},
\qquad
D=D_{\CO,P},
\]
and let $\hat\psi_c$ denote the one-step estimator.

\paragraph{Expansion.}
The usual one-step expansion is
\[
\hat\psi_c-\psi_c
=
(\mathbb P_n-P)D
+
(\mathbb P_n-P)(\hat D-D)
+
R_2.
\]
When the influence function is normalized by
$n^{-1}\sum_i\hat h_i$ rather than $E(\hat h)$, the difference is a
product of $(\mathbb P_n-P)\hat h$ and first-order terms and is therefore
$O_p(n^{-1/2})o_p(1)$. By Proposition~3 and Condition~6,
\[
R_2=o_p(n^{-1/2}).
\]

\paragraph{Empirical-process term.}
For fold $k$ with $n_k$ observations, condition on the training data.
Then $\hat D$ is fixed and
\[
(\mathbb P_{n_k}-P)(\hat D-D)
\]
has conditional mean zero and conditional variance
\[
\frac{\|\hat D-D\|_2^2}{n_k}.
\]
Chebyshev's inequality therefore gives
\[
(\mathbb P_{n_k}-P)(\hat D-D)
=
O_p\left(
\|\hat D-D\|_2n^{-1/2}
\right).
\]
Summing over the fixed number of folds, it suffices that
\[
\|\hat D-D\|_2\to0.
\]

\paragraph{$L_2$ convergence.}
By Proposition~2, $|\hat D|$ and $|D|$ are bounded by a common constant
with probability tending to one. At a fixed observation, $\hat D$
depends on the nuisance estimators through
\[
\hat e(X),\qquad
\hat S_a(\cdot\mid X),\qquad
\frac{\hat G_a(v\mid X)}{\hat G_a(w\mid X)},
\quad v,w\le\tau.
\]
Integrals involving $d\hat\Lambda^C/\hat G$ are of the same form after
integration by parts, because
\[
\frac{d\hat\Lambda^C}{\hat G}
=
d(1/\hat G).
\]
On the event that the errors in Condition~5 are below $\eta$ at $X$,
\[
|\hat D-D|
\le
\varpi(\eta)
\]
for a modulus $\varpi(\eta)\to0$. Condition~5 makes the complementary
event asymptotically negligible, so
\[
\hat D(O)-D(O)\to0
\]
in probability. Bounded convergence then gives
\[
\|\hat D-D\|_2\to0.
\]

\paragraph{Normality and variance estimation.}
Because $D$ is bounded, the central limit theorem applies to
$(\mathbb P_n-P)D$. For the variance estimator,
\[
\begin{split}
|\mathbb P_n\hat D^2-PD^2|
\le{}&
\{\mathbb P_n(\hat D-D)^2\}^{1/2}
\{\mathbb P_n(\hat D+D)^2\}^{1/2}\\
&+
|(\mathbb P_n-P)D^2|.
\end{split}
\]
The first factor tends to zero in probability, the second is bounded,
and the final term tends to zero by the law of large numbers. The same
argument gives
\[
\mathbb P_n\hat D\to0.
\]
Thus the plug-in influence-function variance estimator is consistent.
\qed


\subsection{Proof of Theorem 5}

\paragraph{Uniform bound.}
Proposition~2 holds for every $P\in\mathcal P$ with
\[
E_P(h_c)\ge m_0,
\qquad
s_0\ge s_{\min},
\qquad
c\ge c_{\min},
\]
and therefore gives a common bound $K$ for $|D_{\CO}|$.

\paragraph{Uniform normality.}
For independent observations satisfying
\[
|D|\le K,
\qquad
\operatorname{var}(D)=\sigma_P^2\ge v_{\min},
\]
the Berry--Esseen inequality gives
\[
\sup_z
\left|
\Pr_P\left\{
\frac{\sqrt n(\mathbb P_n-P)D}{\sigma_P}\le z
\right\}
-
\Phi(z)
\right|
\le
\frac{
C_{\mathrm{BE}}E|D|^3
}{
\sigma_P^3\sqrt n
}.
\]
Since
\[
E|D|^3
\le
K\,E(D^2)
=
K\sigma_P^2,
\]
the right-hand side is bounded by
\[
\frac{
C_{\mathrm{BE}}K
}{
\sqrt{v_{\min}}\sqrt n
},
\]
uniformly over $\mathcal P$. If Conditions~5 and~6 hold uniformly, the
remainder and empirical-process terms in the proof of Theorem~4 are
$o_p(n^{-1/2})$ uniformly over $\mathcal P$, and
\[
\hat V/\sigma_P^2\to1
\]
uniformly. Uniform continuity of $\Phi$ then gives the asserted uniform
validity of the studentized intervals.

\paragraph{Contrast with overlap weighting.}
Consider the randomized laws of Appendix~A.4, with
$e\equiv1/2$, event hazard $\lambda=0.25$, and $\tau=4$. They satisfy
the conditions defining $\mathcal P$ with constants independent of $b$.
First,
\[
S_a(\tau\mid X)
=
e^{-\lambda\tau}
=
e^{-1}.
\]
Second,
\[
E(h_c)\ge0.2389
\]
by Table~\ref{tab:A-bounded}. Third, the variance has a uniform analytic
lower bound. By Corollary~2, the event-martingale component of
$D_{\CO}$ is orthogonal to the other components, and
\[
v_a
\ge
\sigma_\tau^2
=
\operatorname{var}\{\min(T,\tau)\},
\]
which is independent of $X$ because the event hazard is constant.
Therefore
\[
E(D_{\CO}^2)
\ge
\frac{
E\{h_c^2(v_1/e+v_0/(1-e))\}
}{
E(h_c)^2
}
\ge
4\sigma_\tau^2
\frac{E(h_c^2)}{E(h_c)^2}
\ge
4\sigma_\tau^2,
\]
where the final inequality follows from
$E(h_c^2)\ge E(h_c)^2$.

For $T\sim\operatorname{Exp}(0.25)$ and $\tau=4$,
\[
E\{\min(T,\tau)\}
=
4(1-e^{-1})
=
2.5285,
\]
and
\[
E\{\min(T,\tau)^2\}
=
32(1-2e^{-1})
=
8.4557.
\]
Hence
\[
4\sigma_\tau^2
=
4\left[
8.4557-(2.5285)^2
\right]
=
8.25,
\]
uniformly in $b$. In contrast,
$E(D_{\OW}^2)$ increases without bound with $b$, as shown in
Table~\ref{tab:A-bounded}. \qed
\end{appendix}

\section{Appendix C: Numerical verification of the theory}

This appendix provides numerical checks of the principal derivations in
Sections~4--5 and Appendix~B. These calculations verify the algebra and
are separate from the finite-sample experiments in Section~6. Unless stated
otherwise, $X\in\{0,1,2\}$ with probabilities $(0.3,0.4,0.3)$,
\[
\operatorname{logit}e(X)=-0.3+0.6X,
\]
event hazard
\[
\lambda_a^T(t\mid X)
=
0.2\exp(0.3X-0.4a+0.2aX),
\]
censoring hazard
\[
\lambda_a^C(t\mid X)
=
0.1\exp(\eta X+0.4a),
\]
and $\tau=3$.

\subsection{Influence function along exact submodels}

We first checked the efficient influence function by perturbing each component
of the observed-data law. With $\eta=0.8$ and $c=0.3$, the submodels perturb
the covariate law, propensity score, event hazard, and censoring hazard,
respectively:
\[
p_t(x)\propto p(x)\{1+tb(x)\},\qquad
\operatorname{logit}e_t(X)=\operatorname{logit}e(X)+ta(X),
\]
with the event and censoring hazards multiplied by
$\exp\{tf(X,A)\}$ and $\exp\{tg(X,A)\}$.

For each direction, the pathwise derivative $d\psi_c/dt$ was evaluated by
central differences and compared with $E(D_{\CO}s)$. Where available, the
latter was also computed in closed form. Monte Carlo calculations used
$2\times10^6$ observations. The Monte Carlo mean of $D_{\CO}$ was
$0.00008$ (SE $0.00168$). Table~\ref{tab:C-pathwise} reports the results.

\begin{table}[h]
\caption{Pathwise derivative of $\psi_c$ and $E(D_{\CO}s)$ along exact
submodels.}
\label{tab:C-pathwise}
\centering\small
\begin{tabular}{lrrrr}
\toprule
Direction
& $d\psi_c/dt$
& Closed form
& Monte Carlo (SE)
& $z$\\
\midrule
Covariate law
& 0.0107915 & 0.0107915 & 0.01290 (0.00122) & 1.73\\
Propensity score
& 0.0105721 & 0.0105721 & 0.00917 (0.00060) & $-$2.33\\
Event hazard
& 0.29575 & -- & 0.29565 (0.00091) & $-$0.11\\
Censoring hazard
& 0.0361144 & 0.0361144 & 0.03483 (0.00077) & $-$1.68\\
\bottomrule
\end{tabular}
\end{table}

The available closed-form calculations agree with the exact derivatives to
within $10^{-11}$. Table~\ref{tab:C-vardec} reports the four components of the orthogonal variance decomposition in Corollary~2.

\begin{table}[h]
\caption{Variance decomposition of Corollary~2: analytic values and Monte
Carlo second moments based on $2\times10^6$ observations.}
\label{tab:C-vardec}
\centering\small
\begin{tabular}{lrr}
\toprule
Component & Analytic & Monte Carlo\\
\midrule
Covariate ($D_X$) & 0.00428 & 0.00428\\
Treatment score ($D_A$) & 0.00105 & 0.00105\\
Event martingale ($D_T$) & 5.60575 & 5.60174\\
Censoring martingale ($D_C$) & 0.01143 & 0.01137\\
Total & 5.62252 & 5.61734\\
\bottomrule
\end{tabular}
\end{table}

The largest absolute empirical cross moment between components was
$0.00051$.

\subsection{Tilt derivatives}

Let
\[
H(e,\ell_1,\ell_0)
=
\left\{
\frac{\kappa_c(e^{\ell_1})}{e}
+
\frac{\kappa_c(e^{\ell_0})}{1-e}
\right\}^{-1},
\qquad
u_a=\frac{c}{e_a e^{\ell_a}},
\qquad
F=\frac{1}{H}.
\]
The derivative identities used in Propositions~2--3 were verified
symbolically.We additionally evaluated their bounds at $2\times10^6$ random points (Table~\ref{tab:C-audit}) with
\[
e\in[0.02,0.98],\qquad c\in[0.01,1],
\qquad g_a\ge10^{-300}.
\]

\begin{table}[h]
\caption{Largest observed ratio of each derivative to its corresponding
bound.}
\label{tab:C-audit}
\centering\small
\begin{tabular}{lr}
\toprule
Bound & Largest ratio\\
\midrule
$H\le e_ag_a/c$ & 1.000\\
$H\le\min(e,1-e)$ & 0.980\\
$|H_e|\le1$ & 0.960\\
$|H_e|\le H/\{e(1-e)\}$ & 0.980\\
$|H_{\ell_a}|\le H$ & 1.000\\
$|H_{\ell_a\ell_a}|\le3H$ & 0.333\\
$|H_{e\ell_a}|\le3g_a/(c\epsilon_e)$ & 0.094\\
$|H_{ee}|\le8/\{e(1-e)\}$ & 0.063\\
$H_{g_a}/e_a\le1/c$ & 1.000\\
\bottomrule
\end{tabular}
\end{table}

For numerical stability, the final quantity was evaluated as
$c\{H/(e_ag_a)\}^2$. The $H_{ee}$ comparison was evaluated wherever the
direct double-precision calculation remained finite.

\subsection{Second-order remainder}

We next evaluated $R_2(\widehat P,P)$ by direct integration over the true
observed-data law, independently of the analytic remainder formulas. For the
cell $(X,A)=(2,1)$ with all three nuisance components perturbed, direct
integration and the analytic expressions both gave
\[
r_a=0.03055360,
\qquad
L_a=0.06115849.
\]

Table~\ref{tab:C-rem} examines the order of the full remainder.

\begin{table}[h]
\caption{$R_2/\epsilon^2$ for CO with $c=0.3$ as the perturbation size
$\epsilon$ decreases.}
\label{tab:C-rem}
\centering\small
\begin{tabular}{lrrrr}
\toprule
Perturbed nuisance
& $\epsilon=0.2$
& $0.1$
& $0.05$
& $0.025$\\
\midrule
Propensity score
& $-$0.00362 & $-$0.00359 & $-$0.00357 & $-$0.00356\\
Event hazard
& 0.00000 & 0.00000 & 0.00000 & 0.00000\\
Censoring hazard
& $-$0.07195 & $-$0.05847 & $-$0.05273 & $-$0.05010\\
All three
& $-$0.08876 & $-$0.09491 & $-$0.09791 & $-$0.09935\\
\bottomrule
\end{tabular}
\end{table}

The ratios stabilize as $\epsilon$ decreases, consistent with the
second-order expansion. When only the event model is perturbed, the remainder
is zero, as implied by the product structure in Proposition~3. A censoring
perturbation need not have this property because the censoring law also enters
the CO target.

To examine the cancellation under deteriorating follow-up, we fixed the
relative event- and censoring-hazard perturbations and increased $\eta$, while keeping the propensity score exact. (Table~\ref{tab:C-cancel})

\begin{table}[h]
\caption{Absolute second-order remainder as follow-up deteriorates in the
stratum $X=2$, with fixed relative nuisance perturbations.}
\label{tab:C-cancel}
\centering\small
\begin{tabular}{rrrrr}
\toprule
$\eta$
& $G_1(\tau\mid X=2)$
& OW
& CO, $c=0.1$
& CO, $c=1$\\
\midrule
0.8 & $1.09\times10^{-1}$ & $4.08\times10^{-3}$
    & $8.52\times10^{-4}$ & $1.59\times10^{-4}$\\
1.2 & $7.20\times10^{-3}$ & $1.25\times10^{-2}$
    & $1.29\times10^{-3}$ & $6.50\times10^{-4}$\\
1.6 & $1.70\times10^{-5}$ & $7.08\times10^{-2}$
    & $1.46\times10^{-3}$ & $1.06\times10^{-3}$\\
2.0 & $2.44\times10^{-11}$ & 1.97
    & $9.38\times10^{-4}$ & $8.42\times10^{-4}$\\
2.4 & $2.41\times10^{-24}$ & $6.41\times10^{3}$
    & $8.12\times10^{-4}$ & $5.72\times10^{-4}$\\
\bottomrule
\end{tabular}
\end{table}

\subsection{Bounded influence function}

Finally, we examined the influence function directly as follow-up
deteriorated. For $c=0.1$, $10^5$ observations were generated at each value
of $\eta$. We computed $D_{\OW}$ and $D_{\CO}$ under the true nuisance
functions, together with $D_{\CO,\widehat P}$ under perturbations of the
propensity score and both hazards. Table~\ref{tab:C-bound} reports the
results.

\begin{table}[h]
\caption{Largest sampled absolute influence value, the uniform bound from
Proposition~2, and
$\|D_{\CO,\widehat P}-D_{\CO,P}\|_2$ under nuisance perturbations.
The bound uses $s_0=0.335$.}
\label{tab:C-bound}
\centering\small
\begin{tabular}{rrrrrrrr}
\toprule
$\eta$
& $G_1(\tau\mid X=2)$
& $\max|D_{\OW}|$
& $\max|D_{\CO}|$
& Bound
& \multicolumn{3}{c}{$\|D_{\CO,\widehat P}-D_{\CO,P}\|_2$}\\
\cmidrule(lr){6-8}
&&&&& $\epsilon=.2$ & $.1$ & $.05$\\
\midrule
0.8 & $1.1\times10^{-1}$   & 6.85             & 6.70  & 1129
    & 0.346 & 0.165 & 0.081\\
1.6 & $1.7\times10^{-5}$   & $1.75\times10^3$ & 10.07 & 1680
    & 0.344 & 0.164 & 0.081\\
2.4 & $2.4\times10^{-24}$  & $2.98\times10^3$ & 16.54 & 2742
    & 0.495 & 0.252 & 0.127\\
3.2 & $1.0\times10^{-117}$ & $2.03\times10^3$ & 19.34 & 3198
    & 0.293 & 0.146 & 0.073\\
\bottomrule
\end{tabular}
\end{table}

The sampled OW maximum depends on whether the finite Monte Carlo sample
contains the most weakly followed profiles; its population influence function
is unbounded in this model. In contrast, the CO influence values remain within
the bound of Proposition~2. The $L_2$ distance between
$D_{\CO,\widehat P}$ and $D_{\CO,P}$ decreases with the perturbation size
throughout the sequence.

\section{Appendix D: Additional simulation results}

\subsection{Complete results from the simulation study}
\label{sec:D-complete}

Table~\ref{tab:D-complete} summarizes the prespecified CO sensitivity
path and alternative weighting and truncation procedures. The primary
OW and CO ($c=0.1$) results, including tail behavior, are reported in
Table~\ref{tab:sim} of the main text. Bias and coverage below are
computed relative to each estimator's own target.


\begin{table}[p]
\caption{Sensitivity analyses from the simulation study, with
$n=2000$ and 2000 replicates per configuration. TO and FO denote
treatment and follow-up overlap. Bias is multiplied by 100 on the
restricted mean survival time scale. Cover is coverage (\%) of the
95\% influence-function interval. VR is variance relative to OW.
TV is total-variation distance from the OW target population.}
\label{tab:D-complete}
\centering
\footnotesize
\setlength{\tabcolsep}{6pt}

\begin{tabular}{cllrrrr}
\toprule
$\tau$ & TO/FO & Estimator & Bias & Cover & VR & TV\\
\midrule

1 & Good/Good & CO, $c=0.3$ & 0.03 & 95.3 & 0.99 & 0.01\\
  &           & CO, $c=1$   & 0.03 & 95.5 & 0.97 & 0.02\\
  &           & Product     & 0.03 & 95.5 & 0.95 & 0.03\\
  &           & $G\ge .05$  & 0.04 & 95.2 & 1.00 & 0.00\\
  &           & $G\ge .10$  & 0.04 & 95.2 & 1.00 & 0.00\\
  &           & 1\% floor   & 0.02 & 95.2 & 1.00 & 0.00\\
  &           & 5\% floor   & $-$0.03 & 95.3 & 1.01 & 0.00\\

\addlinespace
1 & Good/Limited & CO, $c=0.3$ & $-$0.01 & 95.0 & 0.86 & 0.05\\
  &              & CO, $c=1$   & $-$0.01 & 95.2 & 0.82 & 0.09\\
  &              & Product     & $-$0.01 & 95.3 & 0.80 & 0.14\\
  &              & $G\ge .05$  & $-$0.19 & 95.2 & 1.90 & 0.00\\
  &              & $G\ge .10$  & $-$0.29 & 95.7 & 1.73 & 0.00\\
  &              & 1\% floor   & $-$0.38 & 95.8 & 1.32 & 0.00\\
  &              & 5\% floor   & $-$0.71 & 93.0 & 1.34 & 0.00\\

\addlinespace
1 & Limited/Good & CO, $c=0.3$ & 0.05 & 95.0 & 0.99 & 0.01\\
  &              & CO, $c=1$   & 0.05 & 94.9 & 0.98 & 0.02\\
  &              & Product     & 0.05 & 94.8 & 0.95 & 0.03\\
  &              & $G\ge .05$  & 0.05 & 95.2 & 1.00 & 0.00\\
  &              & $G\ge .10$  & 0.05 & 95.2 & 1.00 & 0.00\\
  &              & 1\% floor   & 0.03 & 95.1 & 1.00 & 0.00\\
  &              & 5\% floor   & $-$0.04 & 94.3 & 1.01 & 0.00\\

\addlinespace
1 & Limited/Limited & CO, $c=0.3$ & 0.02 & 94.6 & 0.89 & 0.04\\
  &                 & CO, $c=1$   & 0.02 & 94.8 & 0.86 & 0.09\\
  &                 & Product     & 0.02 & 94.8 & 0.83 & 0.14\\
  &                 & $G\ge .05$  & $-$0.20 & 95.1 & 1.95 & 0.00\\
  &                 & $G\ge .10$  & $-$0.30 & 95.0 & 1.74 & 0.00\\
  &                 & 1\% floor   & $-$0.43 & 95.6 & 1.43 & 0.00\\
  &                 & 5\% floor   & $-$0.84 & 94.1 & 1.39 & 0.00\\

\bottomrule
\end{tabular}
\end{table}


\begin{table}[p]
\ContinuedFloat
\caption{Sensitivity analyses from the simulation study
(continued): $\tau=3$.}
\centering
\footnotesize
\setlength{\tabcolsep}{6pt}

\begin{tabular}{cllrrrr}
\toprule
$\tau$ & TO/FO & Estimator & Bias & Cover & VR & TV\\
\midrule

3 & Good/Good & CO, $c=0.3$ & 0.06 & 94.2 & 0.98 & 0.02\\
  &           & CO, $c=1$   & 0.08 & 94.0 & 0.97 & 0.05\\
  &           & Product     & 0.09 & 93.7 & 0.98 & 0.10\\
  &           & $G\ge .05$  & 0.05 & 94.2 & 1.00 & 0.00\\
  &           & $G\ge .10$  & 0.05 & 94.2 & 1.00 & 0.00\\
  &           & 1\% floor   & $-$0.17 & 93.9 & 1.00 & 0.00\\
  &           & 5\% floor   & $-$0.83 & 92.9 & 1.03 & 0.00\\

\addlinespace
3 & Good/Limited & CO, $c=0.3$ & 0.16 & 95.0 & 0.59 & 0.12\\
  &              & CO, $c=1$   & 0.13 & 95.0 & 0.60 & 0.19\\
  &              & Product     & 0.10 & 94.3 & 0.65 & 0.28\\
  &              & $G\ge .05$  & $-$1.38 & 96.0 & 2.19 & 0.00\\
  &              & $G\ge .10$  & $-$2.00 & 95.2 & 1.57 & 0.00\\
  &              & 1\% floor   & $-$1.64 & 97.2 & 1.11 & 0.00\\
  &              & 5\% floor   & $-$2.72 & 95.3 & 0.83 & 0.00\\

\addlinespace
3 & Limited/Good & CO, $c=0.3$ & 0.17 & 95.5 & 0.99 & 0.02\\
  &              & CO, $c=1$   & 0.19 & 95.3 & 0.98 & 0.05\\
  &              & Product     & 0.19 & 95.2 & 0.97 & 0.10\\
  &              & $G\ge .05$  & 0.16 & 95.6 & 1.00 & 0.00\\
  &              & $G\ge .10$  & 0.16 & 95.6 & 1.00 & 0.00\\
  &              & 1\% floor   & $-$0.26 & 95.4 & 0.99 & 0.00\\
  &              & 5\% floor   & $-$1.46 & 93.8 & 0.99 & 0.00\\

\addlinespace
3 & Limited/Limited & CO, $c=0.3$ & 0.10 & 94.5 & 0.51 & 0.12\\
  &                 & CO, $c=1$   & 0.11 & 95.0 & 0.51 & 0.19\\
  &                 & Product     & 0.13 & 95.2 & 0.53 & 0.28\\
  &                 & $G\ge .05$  & $-$1.90 & 96.2 & 2.02 & 0.00\\
  &                 & $G\ge .10$  & $-$2.40 & 96.0 & 1.44 & 0.00\\
  &                 & 1\% floor   & $-$2.80 & 95.9 & 1.07 & 0.00\\
  &                 & 5\% floor   & $-$4.32 & 94.2 & 0.77 & 0.00\\

\bottomrule
\end{tabular}
\end{table}


\begin{table}[p]
\ContinuedFloat
\caption{Sensitivity analyses from the simulation study
(continued): $\tau=5$.}
\centering
\footnotesize
\setlength{\tabcolsep}{6pt}

\begin{tabular}{cllrrrr}
\toprule
$\tau$ & TO/FO & Estimator & Bias & Cover & VR & TV\\
\midrule

5 & Good/Good & CO, $c=0.3$ & 0.08 & 95.3 & 1.00 & 0.04\\
  &           & CO, $c=1$   & 0.08 & 95.5 & 1.01 & 0.08\\
  &           & Product     & 0.06 & 95.5 & 1.09 & 0.16\\
  &           & $G\ge .05$  & 0.04 & 95.8 & 1.00 & 0.00\\
  &           & $G\ge .10$  & 0.03 & 95.8 & 1.00 & 0.00\\
  &           & 1\% floor   & $-$0.76 & 95.0 & 1.01 & 0.00\\
  &           & 5\% floor   & $-$2.52 & 91.8 & 1.05 & 0.00\\

\addlinespace
5 & Good/Limited & CO, $c=0.3$ & $-$0.16 & 95.3 & 0.11 & 0.18\\
  &              & CO, $c=1$   & $-$0.24 & 95.0 & 0.12 & 0.25\\
  &              & Product     & $-$0.34 & 95.0 & 0.13 & 0.36\\
  &              & $G\ge .05$  & $-$3.15 & 96.0 & 0.41 & 0.00\\
  &              & $G\ge .10$  & $-$4.12 & 96.0 & 0.27 & 0.00\\
  &              & 1\% floor   & $-$2.98 & 97.5 & 0.20 & 0.00\\
  &              & 5\% floor   & $-$4.47 & 97.0 & 0.13 & 0.00\\

\addlinespace
5 & Limited/Good & CO, $c=0.3$ & 0.05 & 94.7 & 1.00 & 0.04\\
  &              & CO, $c=1$   & 0.03 & 95.0 & 1.02 & 0.09\\
  &              & Product     & $-$0.02 & 95.1 & 1.08 & 0.16\\
  &              & $G\ge .05$  & 0.05 & 95.0 & 1.00 & 0.00\\
  &              & $G\ge .10$  & 0.03 & 95.0 & 1.00 & 0.00\\
  &              & 1\% floor   & $-$1.91 & 94.8 & 1.01 & 0.00\\
  &              & 5\% floor   & $-$5.90 & 90.3 & 1.00 & 0.00\\

\addlinespace
5 & Limited/Limited & CO, $c=0.3$ & 0.13 & 94.0 & 0.41 & 0.18\\
  &                 & CO, $c=1$   & 0.10 & 94.3 & 0.42 & 0.25\\
  &                 & Product     & 0.07 & 94.6 & 0.47 & 0.36\\
  &                 & $G\ge .05$  & $-$2.89 & 96.2 & 1.42 & 0.00\\
  &                 & $G\ge .10$  & $-$3.82 & 96.2 & 0.93 & 0.00\\
  &                 & 1\% floor   & $-$4.73 & 96.2 & 0.75 & 0.00\\
  &                 & 5\% floor   & $-$7.83 & 94.5 & 0.52 & 0.00\\

\bottomrule
\end{tabular}
\end{table}

\subsection{Constant-target experiment}

We repeated the limited-follow-up settings with event hazard
\[
0.2\exp\{0.8\tanh(X_3)+A(-0.3+0.4\tanh X_3)\}.
\]
Because $X_3$ is independent of the covariates determining treatment and
follow-up, all tilts have the same estimand: $0.116$ at $\tau=3$ and $0.277$
at $\tau=5$. Thus differences in sampling variance cannot be attributed to
differences in the target effect. Each setting used 500 samples of size $n=2000$; Table~\ref{tab:D-const} reports the results.

\begin{table}[h]
\caption{Constant-target experiment. VR denotes variance relative to OW and
is computed from the reported standard deviations. Truncated OW is evaluated
against the untruncated OW target.}
\label{tab:D-const}
\centering\footnotesize
\begin{tabular}{lrrrrrr}
\toprule
Estimator & Bias & SD & VR & Cover & Kurt & TV\\
\midrule
\multicolumn{7}{l}{\emph{$\tau=3$; treatment overlap good}}\\
OW                  & $-$0.002 & 0.057 & 1.00 & 95.6 & 6.6 & 0.00\\
Product             & $-$0.001 & 0.052 & 0.83 & 95.2 & 2.9 & 0.28\\
CO, $c=0.1$         & $-$0.001 & 0.046 & 0.65 & 95.4 & 2.8 & 0.08\\
CO, $c=0.3$         & $-$0.001 & 0.047 & 0.68 & 95.0 & 2.8 & 0.12\\
CO, $c=1$           & $-$0.001 & 0.048 & 0.71 & 95.0 & 2.9 & 0.19\\
OW, floor 5th pct.  & $-$0.063 & 0.062 & 1.18 & 90.8 & 3.0 & 0.00\\
\addlinespace

\multicolumn{7}{l}{\emph{$\tau=3$; treatment overlap limited}}\\
OW                  & 0.000 & 0.105 & 1.00 & 95.2 & 113.3 & 0.00\\
Product             & 0.001 & 0.070 & 0.44 & 95.6 & 2.8 & 0.28\\
CO, $c=0.1$         & 0.003 & 0.062 & 0.35 & 95.0 & 2.9 & 0.08\\
CO, $c=0.3$         & 0.002 & 0.062 & 0.35 & 95.4 & 2.9 & 0.12\\
CO, $c=1$           & 0.002 & 0.065 & 0.38 & 96.4 & 2.8 & 0.19\\
OW, floor 5th pct.  & $-$0.060 & 0.092 & 0.77 & 93.4 & 3.5 & 0.00\\
\addlinespace

\multicolumn{7}{l}{\emph{$\tau=5$; treatment overlap good}}\\
OW                  & $-$0.003 & 0.123 & 1.00 & 94.8 & 5.2 & 0.00\\
Product             & $-$0.004 & 0.098 & 0.63 & 95.6 & 2.8 & 0.36\\
CO, $c=0.1$         & $-$0.001 & 0.085 & 0.48 & 95.4 & 2.8 & 0.13\\
CO, $c=0.3$         & $-$0.002 & 0.086 & 0.49 & 94.2 & 2.8 & 0.18\\
CO, $c=1$           & $-$0.003 & 0.090 & 0.54 & 94.8 & 2.8 & 0.25\\
OW, floor 5th pct.  & $-$0.146 & 0.115 & 0.87 & 88.2 & 2.9 & 0.00\\
\addlinespace

\multicolumn{7}{l}{\emph{$\tau=5$; treatment overlap limited}}\\
OW                  & $-$0.008 & 0.194 & 1.00 & 96.4 & 48.6 & 0.00\\
Product             & $-$0.003 & 0.131 & 0.46 & 96.0 & 2.8 & 0.36\\
CO, $c=0.1$         & 0.000 & 0.111 & 0.33 & 96.2 & 3.0 & 0.13\\
CO, $c=0.3$         & $-$0.001 & 0.113 & 0.34 & 96.8 & 2.9 & 0.18\\
CO, $c=1$           & $-$0.002 & 0.119 & 0.38 & 96.6 & 2.8 & 0.25\\
OW, floor 5th pct.  & $-$0.146 & 0.179 & 0.85 & 90.6 & 3.1 & 0.00\\
\bottomrule
\end{tabular}
\end{table}

CO remains more stable than OW when all tilts share the same estimand. The
5th-percentile truncation changes a different object: it modifies the
estimator while retaining the OW target, producing appreciable bias in these
settings.

\subsection{Misjudged follow-up}

We next examined a censoring model that fails to identify part of the
weak-follow-up region. Treatment was randomized, follow-up overlap was
limited, and $\tau=5$. The fitted censoring model capped $X_2$ at $k$, thereby
treating follow-up as constant above the threshold. Each setting used 500 samples of size $n=2000$; Table~\ref{tab:D-misjudged}
reports the results.

\begin{table}[h]
\caption{Performance when the fitted censoring model is flat above $X_2=k$.
Mass is $\Pr(X_2>k)$.}
\label{tab:D-misjudged}
\centering\footnotesize
\begin{tabular}{rrrrrrrrrrr}
\toprule
&& \multicolumn{3}{c}{OW}
& \multicolumn{3}{c}{CO, $c=0.1$}
& \multicolumn{3}{c}{CO, $c=1$}\\
\cmidrule(lr){3-5}\cmidrule(lr){6-8}\cmidrule(lr){9-11}
$k$ & Mass
& Bias & SD & Cover
& Bias & SD & Cover
& Bias & SD & Cover\\
\midrule
none & 0.000 & $-$0.002 & 0.120 & 94.0
     & 0.003 & 0.090 & 93.6
     & 0.002 & 0.091 & 95.0\\
2.0  & 0.023 & 0.002 & 0.115 & 93.8
     & 0.003 & 0.090 & 94.2
     & 0.002 & 0.091 & 95.0\\
1.5  & 0.067 & 0.003 & 0.100 & 92.8
     & 0.001 & 0.091 & 94.0
     & 0.002 & 0.091 & 95.4\\
1.0  & 0.159 & 0.003 & 0.094 & 93.2
     & $-$0.008 & 0.090 & 93.4
     & 0.002 & 0.091 & 95.8\\
0.5  & 0.309 & 0.002 & 0.092 & 92.0
     & $-$0.028 & 0.088 & 92.0
     & 0.002 & 0.089 & 96.0\\
\bottomrule
\end{tabular}
\end{table}

For $c=0.1$, bias becomes visible as the probability of the misjudged region
increases, while the sampling variance remains controlled. This experiment
illustrates that CO can only respond to weak follow-up represented by the
fitted censoring law.

\subsection{Alignment of treatment and follow-up information}

To vary the location of weak follow-up relative to treatment information, we
used
\[
\operatorname{logit}e(X)=-1.5+X_2,
\qquad
\lambda^C(t\mid X,A)
=
\lambda_{C0}\exp(1.4\rho X_2+0.3A),
\]
with $\rho\in\{-1,-0.5,0,0.5,1\}$. The outcome depends only on $X_3$, so
all tilts again have the same estimand, $0.277$ at $\tau=5$. Each setting used 500 samples of size $n=2000$; Table~\ref{tab:D-alignment}
reports the results.

\begin{table}[h]
\caption{Alignment of treatment and follow-up information. Weak FU is the
proportion of the OW population with
$\min_aG_a(\tau\mid X)<0.10$; VR is variance relative to OW; ESS is effective
sample size relative to OW.}
\label{tab:D-alignment}
\centering\footnotesize
\begin{tabular}{rrrlrrrrrr}
\toprule
$\rho$ & Weak FU & $G_{1,0.05}$ & Estimator
& VR & ESS & TV & Bias & Cover & Kurt\\
\midrule
$-$1.0 & 0.06 & 0.002 & OW          & 1.00 & 1.00 & 0.00 & 0.002 & 96.0 & 3.2\\
       &&& CO, $c=0.1$              & 0.82 & 0.90 & 0.07 & 0.004 & 96.6 & 3.0\\
       &&& CO, $c=0.3$              & 0.81 & 0.84 & 0.11 & 0.005 & 97.4 & 3.0\\
       &&& CO, $c=1$                & 0.84 & 0.73 & 0.18 & 0.006 & 97.0 & 3.0\\
       &&& Product                  & 0.93 & 0.63 & 0.25 & 0.007 & 96.8 & 2.9\\
\addlinespace
$-$0.5 & 0.01 & 0.077 & OW          & 1.00 & 1.00 & 0.00 & 0.006 & 94.6 & 2.9\\
       &&& CO, $c=0.1$              & 0.99 & 0.94 & 0.04 & 0.007 & 94.0 & 3.1\\
       &&& CO, $c=0.3$              & 1.00 & 0.88 & 0.08 & 0.007 & 94.4 & 3.1\\
       &&& CO, $c=1$                & 1.05 & 0.77 & 0.14 & 0.007 & 95.0 & 3.2\\
       &&& Product                  & 1.18 & 0.65 & 0.22 & 0.007 & 93.8 & 3.3\\
\addlinespace
0.0 & 0.00 & 0.405 & OW             & 1.00 & 1.00 & 0.00 & $-$0.006 & 94.4 & 3.3\\
    &&& CO, $c=0.1$                 & 1.00 & 1.00 & 0.00 & $-$0.006 & 94.8 & 3.3\\
    &&& CO, $c=0.3$                 & 1.00 & 0.99 & 0.01 & $-$0.006 & 94.8 & 3.4\\
    &&& CO, $c=1$                   & 1.03 & 0.98 & 0.01 & $-$0.007 & 94.4 & 3.4\\
    &&& Product                     & 1.06 & 1.00 & 0.00 & $-$0.007 & 94.0 & 3.4\\
\addlinespace
0.5 & 0.10 & 0.082 & OW             & 1.00 & 1.00 & 0.00 & $-$0.001 & 94.2 & 3.2\\
    &&& CO, $c=0.1$                 & 0.95 & 1.09 & 0.07 & $-$0.003 & 94.6 & 3.3\\
    &&& CO, $c=0.3$                 & 1.00 & 1.13 & 0.13 & $-$0.003 & 94.6 & 3.3\\
    &&& CO, $c=1$                   & 1.13 & 1.17 & 0.20 & $-$0.004 & 94.2 & 3.2\\
    &&& Product                     & 1.52 & 1.11 & 0.34 & $-$0.003 & 94.0 & 3.0\\
\addlinespace
1.0 & 0.28 & 0.002 & OW             & 1.00 & 1.00 & 0.00 & $-$0.056 & 95.8 & 478.4\\
    &&& CO, $c=0.1$                 & 0.01 & 1.06 & 0.19 & $-$0.004 & 95.2 & 2.8\\
    &&& CO, $c=0.3$                 & 0.01 & 1.07 & 0.26 & $-$0.005 & 94.6 & 2.8\\
    &&& CO, $c=1$                   & 0.01 & 1.06 & 0.34 & $-$0.006 & 95.6 & 2.9\\
    &&& Product                     & 0.01 & 0.95 & 0.46 & $-$0.007 & 95.2 & 3.1\\
\bottomrule
\end{tabular}
\end{table}

When weak follow-up falls where treatment is rare ($\rho<0$), the CO
tilts perform similarly to OW or modestly improve precision. When weak
follow-up falls where treatment information is concentrated ($\rho>0$),
stronger protection can cost precision at moderate alignment, whereas
under strong alignment OW becomes unstable and all follow-up-aware tilts
substantially reduce variance. Thus, performance depends on the joint
geometry of treatment and follow-up support rather than on either
mechanism alone.

We also considered a rare-treatment design with
$\operatorname{logit}e(X)=-2.8+X_2$, corresponding to approximately
$8\%$ treated. Table~\ref{tab:rare_alignment} reports the results.

\begin{table}[h]
\caption{Alignment with rare treatment,
$\operatorname{logit}e(X)=-2.8+X_2$ (about $8\%$ treated), based on
400 samples of size $n=2000$ per value of $\rho$.
IQR$^2$ is the squared ratio of interquartile ranges relative to OW,
providing a tail-insensitive precision comparison.}
\label{tab:rare_alignment}
\centering\footnotesize
\begin{tabular}{rrrlrrrr}
\toprule
$\rho$ & Weak FU & $G_{1,0.05}$ & Estimator & VR & IQR$^2$ & Kurt & TV\\
\midrule
0.50 & 0.21 & 0.072 & OW & 1.00 & 1.00 & 3.3 & 0.00\\
 & & & CO, $c=0.1$ & 0.94 & 0.99 & 3.1 & 0.12\\
 & & & CO, $c=0.3$ & 1.01 & 1.03 & 3.3 & 0.19\\
 & & & CO, $c=1$ & 1.17 & 1.08 & 3.2 & 0.27\\
 & & & Product & 1.62 & 1.70 & 3.0 & 0.41\\
\addlinespace
0.75 & 0.35 & 0.016 & OW & 1.00 & 1.00 & 15.4 & 0.00\\
 & & & CO, $c=0.1$ & 0.53 & 0.90 & 5.2 & 0.22\\
 & & & CO, $c=0.3$ & 0.62 & 0.92 & 19.5 & 0.30\\
 & & & CO, $c=1$ & 0.71 & 0.99 & 20.5 & 0.38\\
 & & & Product & 0.75 & 1.24 & 3.0 & 0.50\\
\bottomrule
\end{tabular}
\end{table}

\subsection{Sensitivity to the censoring model}

We generated censoring from
\[
\lambda^C(t\mid X,A)
=
\lambda_{C0}
\exp\{0.3X_2+1.5\max(X_2-0.5,0)+0.3A\},
\]
which produces weak follow-up in the upper tail of $X_2$. The weak-follow-up
proportion of the OW population is $0.12$, the fifth percentile of
$G_1(\tau\mid X)$ is $0.006$, and all tilts have estimand $0.277$ at
$\tau=5$. The censoring law was fitted correctly, linearly in $X_2$, or with
$X_2$ capped at $0.5$. The same 500 samples of size $n=2000$ were used for all three fits (Table~\ref{tab:D-censmodel}).

\begin{table}[h]
\caption{Sensitivity to censoring-model specification. SD is reported under
the correct censoring model; mean absolute change compares each alternative
fit with the correct fit on the same simulated samples.}
\label{tab:D-censmodel}
\centering\footnotesize
\begin{tabular}{lrrrrrr}
\toprule
& \multicolumn{3}{c}{Bias}
&& \multicolumn{2}{c}{Mean $|$change$|$}\\
\cmidrule(lr){2-4}\cmidrule(lr){6-7}
Estimator
& Correct & Linear & Flat tail
& SD
& Linear & Flat tail\\
\midrule
OW          & 0.005 & $-$0.003 & $-$0.002 & 0.248 & 0.049 & 0.054\\
CO, $c=0.1$ & $-$0.004 & $-$0.004 & $-$0.003 & 0.087 & 0.005 & 0.011\\
CO, $c=0.3$ & $-$0.005 & $-$0.005 & $-$0.004 & 0.087 & 0.005 & 0.011\\
CO, $c=1$   & $-$0.006 & $-$0.007 & $-$0.006 & 0.089 & 0.009 & 0.013\\
Product     & $-$0.007 & $-$0.010 & $-$0.009 & 0.092 & 0.020 & 0.020\\
\bottomrule
\end{tabular}
\end{table}

In this experiment, changing the censoring specification moved the CO
estimates less than the OW estimate. This is an empirical sensitivity result,
not a robustness property of the estimand.

\section{Appendix E: Additional Rotterdam results}

\subsection{Follow-up pattern}

The Rotterdam cohort contains 2982 women, of whom 1272 died and 1710 were
censored. Among censored women, the correlation between year of surgery and
observed follow-up was $-0.78$. Because only the year of surgery is recorded,
we approximated calendar time of last follow-up by
\[
\text{year of surgery}+0.5+\text{follow-up time}.
\]
On this scale, censored follow-up ended at a median of 1998.3, with 10th and
90th percentiles 1995.7 and 2000.3. Only four deaths occurred after the
latter point. Hormonal therapy use increased over calendar time: 2.0\%,
4.2\%, and 16.1\% among women operated on in 1978--1982, 1983--1987, and
1988--1993, respectively.

\subsection{Design diagnostics}

Table~\ref{tab:E-design} reports the follow-up diagnostics under the primary
arm-specific Cox censoring model and the year-stratified Kaplan--Meier
sensitivity model. These quantities were computed at the design stage.

\FloatBarrier
\begin{sidewaystable}[!ht]
\caption{Rotterdam design diagnostics. Equipoise denotes women with
$0.2<\widehat e(X)<0.8$ ($n=469$ for hormonal therapy and $n=775$ for
chemotherapy). Mass is target mass relative to OW. Under year-stratified
Kaplan--Meier estimation, $\widehat G_1=\widehat G_0$ by construction.
The proportions of estimated propensity scores in $(0.1,0.9)$ were 28\%
and 47\% for hormonal therapy and chemotherapy, respectively.}
\label{tab:E-design}

\centering
\footnotesize
\begin{tabular}{llrrrrrrrrr}
\toprule
&&&&&
\multicolumn{2}{c}{Equipoise, \%}
& \multicolumn{3}{c}{Mass}
&\\
\cmidrule(lr){6-7}\cmidrule(lr){8-10}
Treatment & Censoring model & $\tau$
& $\widehat G_{1,.05}$ & $\widehat G_{0,.05}$
& $\widehat G_1<.10$ & $\widehat G_1<.05$
& CO .1 & CO 1 & Product
& TV (CO .1)\\
\midrule
Hormonal & Cox & 5 & .709 & .764 & 0.0 & 0.0 & .99 & .89 & .80 & .01\\
Hormonal & Cox & 7 & .199 & .337 & 9.6 & 2.6 & .90 & .64 & .51 & .05\\
Hormonal & Cox & 8 & .021 & .084 & 34.3 & 27.5 & .73 & .46 & .33 & .18\\
\addlinespace
Hormonal & KM by year & 5 & .704 & .704 & 0.0 & 0.0 & .99 & .90 & .82 & .01\\
Hormonal & KM by year & 7 & .129 & .129 & 0.0 & 0.0 & .92 & .70 & .56 & .05\\
Hormonal & KM by year & 8 & .037 & .037 & 36.9 & 36.9 & .76 & .52 & .40 & .18\\
\addlinespace
Chemotherapy & Cox & 5 & .787 & .756 & 0.0 & 0.0 & .99 & .95 & .90 & .00\\
Chemotherapy & Cox & 7 & .290 & .334 & 0.0 & 0.0 & .96 & .80 & .69 & .02\\
Chemotherapy & Cox & 8 & .100 & .078 & 8.4 & 0.1 & .90 & .68 & .56 & .06\\
\addlinespace
Chemotherapy & KM by year & 5 & .704 & .704 & 0.0 & 0.0 & .99 & .94 & .89 & .00\\
Chemotherapy & KM by year & 7 & .129 & .129 & 0.0 & 0.0 & .96 & .81 & .71 & .03\\
Chemotherapy & KM by year & 8 & .037 & .037 & 17.2 & 17.2 & .86 & .68 & .58 & .10\\
\bottomrule
\end{tabular}
\end{sidewaystable}

\subsection{Censoring-model sensitivity}

Table~\ref{tab:E-horm-KM} reports the hormonal-therapy analysis under the
year-stratified Kaplan--Meier censoring model. Kurt(IF) is the kurtosis of
the estimated influence function, and Top 10 is the percentage of its
variance contributed by the ten observations with the largest absolute
influence values.

\begin{table}[h]
\caption{Hormonal therapy under the year-stratified Kaplan--Meier censoring
model.}
\label{tab:E-horm-KM}
\centering\footnotesize
\begin{tabular}{llrrlrrrr}
\toprule
$\tau$ & Target & $\widehat\psi$ & SE & 95\% interval
& Mass & TV & Kurt(IF) & Top 10\\
\midrule
5 & OW          & .262 & .076 & (.114,.410) & 1.00 & .00 & 35.0 & 27.6\\
  & CO, $c=.1$  & .263 & .076 & (.114,.412) & .99 & .01 & 35.4 & 27.7\\
  & CO, $c=.3$  & .265 & .077 & (.114,.415) & .96 & .02 & 36.1 & 28.1\\
  & CO, $c=1$   & .269 & .080 & (.114,.425) & .90 & .05 & 38.3 & 28.9\\
  & Product     & .276 & .084 & (.112,.440) & .82 & .09 & 41.1 & 29.7\\
\addlinespace
7 & OW          & .388 & .123 & (.146,.630) & 1.00 & .00 & 24.0 & 20.3\\
  & CO, $c=.1$  & .402 & .129 & (.149,.656) & .92 & .05 & 26.0 & 21.5\\
  & CO, $c=.3$  & .418 & .136 & (.152,.685) & .84 & .09 & 28.0 & 22.7\\
  & CO, $c=1$   & .450 & .148 & (.161,.740) & .70 & .16 & 32.0 & 25.2\\
  & Product     & .491 & .164 & (.169,.814) & .56 & .25 & 38.1 & 28.4\\
\addlinespace
8 & OW          & .259 & .193 & ($-.119$,.638) & 1.00 & .00 & 315.3 & 45.4\\
  & CO, $c=.1$  & .409 & .182 & (.052,.765) & .76 & .17 & 32.2 & 23.6\\
  & CO, $c=.3$  & .467 & .191 & (.092,.842) & .66 & .23 & 29.2 & 22.2\\
  & CO, $c=1$   & .533 & .207 & (.128,.939) & .52 & .31 & 35.0 & 25.3\\
  & Product     & .608 & .230 & (.157,1.058) & .40 & .38 & 47.1 & 30.7\\
\bottomrule
\end{tabular}
\end{table}

For chemotherapy, follow-up deterioration was less pronounced under the
primary Cox model. Table~\ref{tab:E-chemo8} therefore reports the 8-year
comparison, where the censoring-model sensitivity is most visible.

\begin{table}[h]
\caption{Chemotherapy at $\tau=8$ under the primary Cox and year-stratified
Kaplan--Meier censoring models.}
\label{tab:E-chemo8}
\centering\footnotesize
\begin{tabular}{llrrlrrrr}
\toprule
Model & Target & $\widehat\psi$ & SE & 95\% interval
& Mass & TV & Kurt(IF) & Top 10\\
\midrule
Cox & OW         & .441 & .129 & (.187,.694) & 1.00 & .00 & 15.9 & 14.3\\
    & CO, $c=.1$ & .455 & .135 & (.191,.719) & .90 & .06 & 16.4 & 14.3\\
    & CO, $c=.3$ & .459 & .139 & (.187,.732) & .82 & .11 & 16.7 & 13.9\\
    & CO, $c=1$  & .464 & .145 & (.181,.748) & .68 & .18 & 18.0 & 15.3\\
    & Product    & .473 & .149 & (.181,.765) & .56 & .26 & 20.0 & 17.6\\
\addlinespace
KM  & OW         & .378 & .148 & (.088,.669) & 1.00 & .00 & 111.0 & 28.5\\
    & CO, $c=.1$ & .417 & .142 & (.139,.694) & .86 & .10 & 17.5 & 15.2\\
    & CO, $c=.3$ & .434 & .143 & (.153,.714) & .79 & .15 & 16.7 & 13.8\\
    & CO, $c=1$  & .460 & .145 & (.175,.744) & .68 & .20 & 17.4 & 14.6\\
    & Product    & .494 & .149 & (.202,.786) & .58 & .27 & 20.0 & 17.1\\
\bottomrule
\end{tabular}
\end{table}

At eight years, changing the censoring specification shifted the hormonal
therapy estimate by $0.169$ under OW and $0.098$ under CO with $c=0.1$.
For chemotherapy, the corresponding shifts were $0.062$ and $0.038$.

\subsection{Variance decomposition}

Table~\ref{tab:E-vardec} decomposes the plug-in variance of the primary
hormonal-therapy estimates according to Corollary~2.

\begin{table}[h]
\caption{Variance decomposition for the primary hormonal-therapy analysis.
Cross is the percentage of total variance attributable to cross terms.}
\label{tab:E-vardec}
\centering\footnotesize
\begin{tabular}{llrrrrrr}
\toprule
$\tau$ & Target & $D_X$ & $D_A$ & $D_T$ & $D_C$ & Total & Cross\\
\midrule
5 & OW         & .00006 & .00012 & .00556 & .00000 & .00571 & $-.56$\\
5 & CO, $c=.1$ & .00006 & .00013 & .00563 & .00000 & .00578 & $-.60$\\
7 & OW         & .00016 & .00041 & .01483 & .00000 & .01542 & .10\\
7 & CO, $c=.1$ & .00019 & .00049 & .01639 & .00002 & .01721 & .66\\
8 & OW         & .00023 & .00061 & .02155 & .00000 & .02240 & .00\\
8 & CO, $c=.1$ & .00036 & .00110 & .03067 & .00004 & .03230 & .39\\
\bottomrule
\end{tabular}
\end{table}

At eight years, the event-martingale component increases from $0.02155$
under OW to $0.03067$ under CO, whereas the censoring component under CO is
$0.00004$. Thus the higher primary-analysis variance under CO arises
predominantly from the reweighting of event information rather than from
estimation of the censoring-dependent target.

\subsection{Bootstrap precision comparison}

The subject-level bootstrap was specified in a protocol addendum before it
was run. Of 1000 resamples, 52 failed to converge in a Cox fit and 21
produced missing estimates, leaving 927 usable resamples. Failed and usable
resamples had similar numbers of hormonally treated women (medians 335 and
339) and treated women operated on during 1992--1993 (128 in both groups).

For each usable resample, let $R_b$ denote the ratio of squared
influence-functionstandard errors for CO with $c=0.1$ relative to OW. Table~\ref{tab:E-boot} summarizes the results.
\begin{table}[h]
\caption{Bootstrap comparison of precision in the hormonal-therapy analysis
(927 usable resamples). Boot VR is the ratio of bootstrap variances of the
CO and OW point estimates. ESS$_1$ is treated-arm effective sample size for
CO relative to OW.}
\label{tab:E-boot}
\centering\footnotesize
\begin{tabular}{rrlrlrrr}
\toprule
$\tau$
& Usable
& Median $R_b$ (2.5\%,97.5\%)
& $P(R_b>1)$
& Median ESS$_1$ ratio
& Boot VR
& Kurt OW
& Kurt CO\\
\midrule
5 & 927 & 1.01 (1.01,1.02) & .998 & .99 (.99,1.00) & 1.01 & 2.8 & 2.8\\
7 & 927 & 1.12 (1.05,1.20) & .991 & .94 (.91,.96) & 1.04 & 2.9 & 2.8\\
8 & 927 & 1.40 (1.06,1.62) & .983 & .78 (.73,.83) & 1.19 & 3.0 & 2.7\\
\bottomrule
\end{tabular}
\end{table}

The precision difference increases with the restriction horizon: the median
squared-SE ratio rises from 1.01 at five years to 1.40 at eight years, while
the treated-arm effective sample-size ratio falls to 0.78. Although the
estimated influence functions in the original sample are concentrated in a
small number of observations, the bootstrap distributions of the point
estimates are not heavy-tailed; at eight years their kurtoses are 3.0 for OW
and 2.7 for CO.

\end{document}